\documentclass[letterpaper,twocolumn,english,aps,prl,longbibliography]{revtex4-1}

\usepackage{amsmath,amssymb,amsthm}
\usepackage{hyperref}
\usepackage{amsthm}
\usepackage{graphicx}
\usepackage{subfigure}
\usepackage{color,bbm}
\usepackage{amsfonts}
\usepackage{mathrsfs}
\usepackage{booktabs}
\usepackage{tabularx}
\usepackage{float}
\usepackage{hhline}
\usepackage{bigints}

\begin{document}
\date{\today}
\title{\bf Dynamical Mean-Field Theory of Complex Systems on Sparse Directed Networks}
\author{Fernando L. Metz}
\affiliation{Physics Institute, Federal University of Rio Grande do Sul, 91501-970 Porto Alegre, Brazil}
\begin{abstract}
  Although real-world complex systems typically interact through sparse and heterogeneous networks, analytic
  solutions of their dynamics
  are limited to models with all-to-all interactions.
  Here, we solve the dynamics of a broad range of nonlinear models of complex systems on sparse directed networks
  with a random structure.
  By generalizing dynamical mean-field theory to sparse systems, we derive
an exact equation
for the path-probability describing the effective
  dynamics of a single degree of freedom.
  Our general solution
  applies to key models in the study of neural networks, ecosystems, epidemic spreading, and synchronization.
  Using the population dynamics algorithm, we solve the path-probability equation to determine 
  the phase diagram of a seminal neural network model in the sparse regime, showing that this model
  undergoes a transition from a fixed-point phase to chaos as a function of the network topology.
 \end{abstract}

\maketitle

\paragraph{Introduction.}

Complex dynamical systems are modeled by
$N$ degrees of freedom $x_i(t)$ ($i=1,\dots,N$) that evolve
in time according to the differential equation 
\begin{equation}
  \dot{x}_i(t) = - f(x_i) + \sum_{j=1}^N A_{ij} g(x_i,x_j),
  \label{uds}
\end{equation}  
where $\{ A_{ij} \}_{i,j=1,\dots,N}$ defines the interaction network.
The function $f(x)$ governs the dynamics
in the absence of interactions, and the kernel $g(x,x^{\prime})$ shapes the pairwise couplings.
Equation (\ref{uds}) models the nonequilibrium dynamics of neural networks \cite{Sompolinsky1988,Kadmon2015,Marti2018,Crisanti2018,Martorell2024} and
ecosystems \cite{Opper1992,Bunin2017,Altieri2021}, epidemic spreading \cite{Mata2013,VanMieghem2012,Goltsev2012,Pastor2015}, synchronization phenomena \cite{Stiller1998,Rodrigues2016,Pruser2024}, opinion
dynamics \cite{Baumann2020,Baumann2021}, and  multivariate Ornstein-Uhlenbeck processes \cite{Godreche2019,Crisanti2023,Ferraro2024}.
Table \ref{TableModels} specifies $f(x)$ and $g(x,x^{\prime})$ for paradigmatic models of complex behavior.

\begin{table}[htbp]
    \centering
    \begin{tabular}{l>{\centering\arraybackslash}m{1.2cm}>{\centering\arraybackslash}m{1.4cm}}
        \hline
        \textbf{Model} & $f(x)$ & $g(x,x^{\prime})$ \\
        \hline
        Ornstein-Uhlenbeck process \cite{Godreche2019} & $x$ & $x^{\prime}$ \\
        SIS model of epidemic spreading \cite{Mata2013} & $x$ & $(1-x) x^{\prime}$ \\
        Lotka-Volterra (LV) model \cite{Bunin2017,Altieri2021} & $\;\, x(x-1)$ & $x x^{\prime}$ \\
        Neural network (NN) model \cite{Sompolinsky1988,Crisanti2018} & $x$ & $\tanh(x^{\prime})$ \\
        Kuramoto model \cite{Rodrigues2016} & 0 & $\sin(x^{\prime}-x)$ \\
        \hline
    \end{tabular}
    \caption{Explicit form of $f(x)$ and $g(x,x^{\prime})$ in complex systems modeled by Eq. (\ref{uds}). Equation (\ref{uds})
        for the SIS model is derived through the quenched mean-field approximation \cite{Pastor2015,BookKiss2017,Silva2020}.}
     \label{TableModels}
\end{table}

The foremost problem in the study of complex systems is how to reduce the dynamics of many
interacting elements to the dynamics of a few variables \cite{Thibeault2024}.
Dynamical mean-field theory (DMFT) \cite{Martin1973,Coolen2001,Hertz2017}
is a powerful method to tackle this problem in the limit $N \rightarrow \infty$, yielding 
a solution in terms of
the path-probability for the effective dynamics of a single degree of freedom.
The application of DMFT to models described by Eq. (\ref{uds}) has been attracting
an enormous interest
\cite{Kadmon2015,Marti2018,Crisanti2018,Roy2019,Martorell2024,Bunin2017,Galla2018,Schuecker2018,Mastrogiuseppe2018,Altieri2021,Pruser2024,Crisanti2023,Poley2023,Ferraro2024,Aguirre2024,Park2024,Tuan2024}, especially in the
context of neural networks and ecosystems.
Phase diagrams of these models reveal a rich phenomenology, including
different types of phase transitions \cite{Bunin2017,Mastrogiuseppe2018,Martorell2024}, chaotic behavior \cite{Sompolinsky1988,Marti2018}, and coexistence of
multiple attractors \cite{Bunin2017,Altieri2021,Ros2023}. Despite this substantial theoretical progress, DMFT is currently
limited to dense networks, where each dynamical variable is essentially coupled to {\it all} others by means of
Gaussian interaction strengths. However, the interactions 
in real-world complex
systems are known to be {\it sparse} and {\it heterogeneous} \cite{Newman2010,Guimaraes2020}.
Sparseness indicates that each element
of the system interacts on average with a finite number of others, while heterogeneity
refers to fluctuations in the local topology of the interaction network.
How to integrate these more realistic features in the formalism of DMFT for systems modeled by Eq. (\ref{uds}) remains an
unresolved challenge, and even basic questions, such as deriving the phase diagram of sparse complex systems, are
still out of reach.

On the other hand, the  dynamics of complex systems on sparse networks 
has been extensively studied in the case of Ising spins
\cite{Hatchett2004,Neri2009,Mimura2009,Roudi2011,Aurell2012,Zhang2012,Ferraro2015,Aurell2017,Vazquez2017,Torrisi2021,Torrisi2022,Behrens2023,Behrens2024}.
In this context, both the cavity
method \cite{Neri2009} and DMFT \cite{Hatchett2004,Mimura2009} provide an analytic solution in terms of the
path-probability for the effective dynamics of a single dynamical variable.
However, the presence of bidirected edges induces a temporal feedback, which, in the particular case
of sparse systems, leads to an exponential growth of the path-probability dimension \cite{Hatchett2004,Mimura2009,Behrens2023}, rendering
numerical computations unfeasible.
Approximation schemes, such as the one-time approximation \cite{Neri2009}, make these computations possible \cite{Zhang2012,Aurell2012,Torrisi2022}.
When the interactions are directed or unidirectional, there is no temporal feedback and
  the path-probability equation can be efficiently solved \cite{Derrida1987,Hatchett2004,Neri2009}.

Inspired by these results for Ising spins, in this Letter we solve the dynamics of models governed by Eq. (\ref{uds}) on sparse
directed networks with an heterogeneous topology.
The solution represents a foundational step in the study of complex systems, as it ultimately incorporates
the sparse and heterogeneous structure of complex networks into the formalism of DMFT. Directed networks
are interesting because they model the non-reciprocal interactions in real-world
complex systems \cite{Asllani2018}, including the human cortex  \cite{Peng2024}, food-webs \cite{Dunne2002,Bascompte2009,Guimaraes2020}, gene
regulatory networks \cite{Shen2002,Lee2002}, online social networks \cite{Kwak2010,Schweimer2022}, and the World Wide
Web \cite{PastorSatorras2004}. The formalism presented here thus opens the possibility to analytically
investigate how realistic interactions impact the dynamics of complex systems.

By generalizing DMFT to sparse systems, we obtain an exact equation for the path-probability
describing the effective dynamics of a single variable for $N \rightarrow \infty$, and we solve this equation for different  models in table \ref{TableModels}
by using the population dynamics algorithm \cite{Mezard2001,Mezard2003,Kuhn2008,Metz2019,AbouChacra1973,MezardBook}.
The excellent agreement between our theoretical results and numerical simulations
of finite systems confirms that  our solution applies to various nonlinear models interacting through different network topologies, including
networks with power-law degree distributions \cite{Newman2010}.

As an application, we determine the phase diagram of the neural network model by Sompolinsky {\it et al} \cite{Sompolinsky1988} in
the sparse regime.
We show that the phase diagram displays trivial and nontrivial
fixed-point phases, a chaotic phase with zero mean activity, and a chaotic phase with nonzero mean
activity \cite{Mastrogiuseppe2018}. By calculating certain macroscopic observables, we determine the
transition lines as functions of the mean degree and the variance of the coupling strengths, showing
their consistency with the universal critical lines derived from random matrix
theory \cite{Neri2016,Neri2020,Metz2021}. In particular, we provide numerical evidence that the
transition between the chaotic phases coincides with the vanishing of the gap between
the leading and the subleading eigenvalue of the interaction matrix.


\paragraph{Sparse directed networks.}

Let $A_{ij} = C_{ij} J_{ij}$ be the elements of the $N \times N$ interaction matrix $\boldsymbol{A}$. The binary variables $C_{ij} \in \{ 0,1 \}$ determine
the network topology, while $J_{ij} \in \mathbb{R}$ controls the interaction strengths.
If $C_{ij}=1$, there is a directed
edge $j \rightarrow i$ pointing from node $j$ to $i$, while $C_{ij}=0$ otherwise. 
The indegree $K_i= \sum_{j=1}^N C_{ij}$ and the outdegree $L_i = \sum_{j=1}^N C_{ji}$ count the number of links
entering and leaving node $i$ \cite{Newman2010}, respectively.
The random variables $\{ C_{ij} \}_{i \neq j}$ follow the distribution
\begin{align}
  \mathbb{P}(\{ C_{ij} \}) &= \frac{1}{\mathcal{N}}  \prod_{i \neq j=1}^N \left[ \frac{c}{N} \delta_{C_{ij},1} + \left(1 - \frac{c}{N} \right) \delta_{C_{ij},0} \right]  \nonumber \\
  &\times \prod_{i=1}^N \delta_{K_i, \sum_{j=1}^N C_{ij} } \delta_{L_i, \sum_{j=1}^N C_{ji} },
  \label{gesd}
\end{align}  
where $\mathcal{N}$ is the normalization constant and $C_{ii}=0$. The degrees $\{ K_i,L_i \}_{i=1,\dots,N}$
are independent and identically distributed random variables drawn from 
$p_{k,\ell} = p_{{\rm in},k} \, p_{{\rm out},\ell}$,
where $p_{{\rm in},k}$ and $p_{{\rm out},\ell}$ are the indegree and the outdegree distribution \cite{Newman2010,Metz2019}, respectively.
The parameter $c$ is the average degree
\begin{equation}
c = \sum_{k=0}^{\infty} k \, p_{{\rm in},k} = \sum_{\ell=0}^{\infty} \ell \, p_{{\rm out},\ell}.
\end{equation}  

Equation (\ref{gesd}) defines a network ensemble where directed links
are randomly placed between pairs of nodes with probability $c/N$, subject to the prescribed degree sequences generated from $p_{k,\ell}$.
In the limit $N \rightarrow \infty$, network samples generated from Eq. (\ref{gesd})
are similar to those produced by the configuration model \cite{Newman2001,Fosdick2018}.
The coupling strengths $\{ J_{ij} \}_{i,j=1,\dots,N}$ are independent and identically
distributed random variables drawn from a distribution $p_J$ with mean $\mu_J$ and variance $\sigma_J^2$.
The distributions $p_{k,\ell}$ and $p_J$ fully specify the network ensemble, allowing for
a systematic investigation of how network heterogeneities impact the dynamics of complex systems.


\paragraph{Solution through DMFT.}

We solve the coupled dynamics of Eq. (\ref{uds}) on directed networks by using dynamical
mean-field theory (DMFT) \cite{Martin1973,Coolen2001,Hertz2017}. We consider the sparse regime, where the mean degree $c$ is {\it finite}, independent of $N$.
DMFT is based on the  generating functional
\begin{equation}
  \mathcal{Z} \left[ \boldsymbol{\psi} \right] = \int \left( \prod\limits_{i=1}^N D x_i   \right) \mathcal{P}[\boldsymbol{x}] \, e^{i \int dt \sum\limits_{i=1}^N x_i(t) \psi_i(t)  }
  \label{juju}
\end{equation}  
of the probability density $\mathcal{P}[\boldsymbol{x}]$  of observing a dynamical path of states $\boldsymbol{x}(t) = (x_1(t),\dots,x_N(t))$ in a fixed time interval.
The correlation functions of $\{ x_i(t) \}_{i=1,\dots,N}$ follow from the derivatives
of $\mathcal{Z} \left[ \boldsymbol{\psi} \right]$ with respect to the external sources $\boldsymbol{\psi}(t) = (\psi_1(t),\dots,\psi_N(t))$. The $n$th-moment of
the local variable $x_i(t)$,
\begin{equation}
  \langle x_i^{n}(t) \rangle = \int \left( \prod\limits_{i=1}^N D x_i \right) x_i^{n}(t)  \mathcal{P}[\boldsymbol{x}]
  = (- i)^n \frac{\delta^n \mathcal{Z} \left[ \boldsymbol{\psi} \right]  }{ \delta \psi_i^n(t)} \Bigg{|}_{\boldsymbol{\psi}=0}, 
\end{equation}  
yields the time-evolution of the macroscopic quantities
\begin{equation}
m(t) = \lim_{N \rightarrow \infty} \frac{1}{N} \sum_{i=1}^N \langle x_i(t) \rangle, \quad  q(t) = \lim_{N \rightarrow \infty} \frac{1}{N} \sum_{i=1}^N \langle x_i^{2}(t) \rangle.
\end{equation}  
Clearly, $\mathcal{Z} \left[ \boldsymbol{\psi} \right]$ fulfills  $\mathcal{Z} \left[0 \right] =1$.

In \footnote{See the Supplemental Material for all derivations of DMFT and a detailed
account of the population dynamics algorithm.}, we calculate the  average of $\mathcal{Z} \left[ \boldsymbol{\psi} \right]$
over the network ensemble defined by Eq. (\ref{gesd}) for finite $c$, recasting the problem in terms of the solution of a saddle-point integral.
More importantly, we give
a clear physical interpretation of the order-parameters, simplifying the saddle-point equations and obtaining a
feasible solution for the dynamics in the sparse regime. In the limit $N \rightarrow \infty$, the microscopic dynamical variables
decouple, and the path-probability $\mathcal{P}[x]$ for the effective dynamics of a single variable $x(t)$
is determined from
\begin{align}
  \mathcal{P}[x] &= \sum\limits_{k=0}^\infty p_{{\rm in},k} \int \left( \prod\limits_{j=1}^k D x_j \mathcal{P}[x_j] \right) \int \left( \prod\limits_{j=1}^k d J_j p_J(J_j) \right)  \nonumber \\
  &\times  \delta_F \Bigg[ \dot{x}(t) + f(x(t)) - \sum\limits_{j=1}^k J_j \, g(x(t),x_j(t) ) \Bigg] ,
  \label{hu1}
\end{align}  
where $\delta_F$ is the functional Dirac-$\delta$.
The macroscopic observables are computed from
$m(t) = \langle x(t) \rangle_{*}$ and $q(t) = \langle x^2(t) \rangle_{*}$, where
\begin{equation}
\langle x^n (t) \rangle_{*} = \int D x \, x^n (t) \mathcal{P}[x] \quad (n=1,2)
\end{equation}  
is the average over the effective dynamics governed by $\mathcal{P}[x]$.
The self-consistent Eq. (\ref{hu1}) is the exact solution of a broad class of models (see table \ref{TableModels})
on sparse directed networks with a local treelike structure \cite{Bordenave2010} and arbitrary distributions $p_J$ and $p_{k,\ell} = p_{{\rm in},k} p_{{\rm out},\ell}$.
In  \cite{Note1}, we address the more general case of dynamical models with Gaussian additive noise on networks with correlated indegrees and outdegrees.
The solution of Eq. (\ref{hu1}) determines the time-evolution of the full probability distribution of the microscopic variables
in the limit $N \rightarrow \infty$.

\begin{figure}[H]
  \begin{center}
    \includegraphics[scale=0.66]{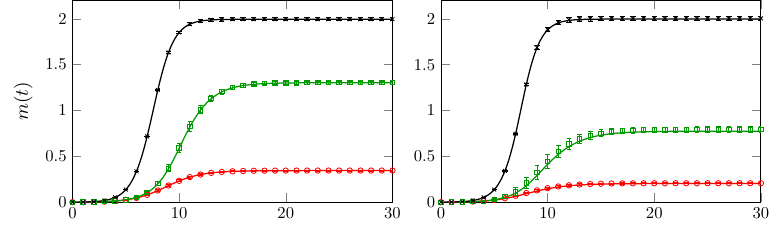} \\
\hspace{-0.2cm}
    \includegraphics[scale=0.66]{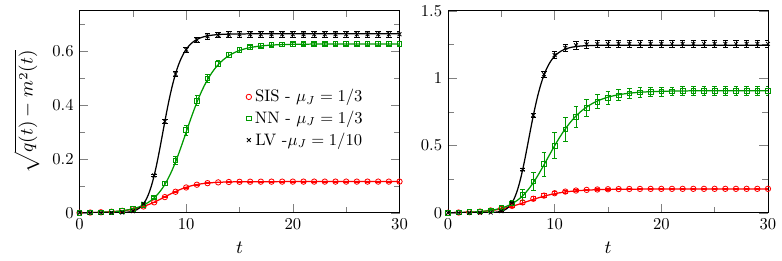}
    \caption{Dynamics of the mean $m(t)$ and the standard deviation $\sqrt{q(t) - m^2(t)}$
      for the SIS model, LV model, and NN
      model (see table \ref{TableModels}).
      Results are shown for directed random networks with average degree $c=5$ and two indegree distributions (Eq. (\ref{degru})): Poisson (left column) and
      geometric (right column). The distribution $p_J$ of the coupling strengths has mean $\mu_J$ and standard deviation $\sigma_J=0.1$. For the
      LV and NN models, $p_J$ is Gaussian; for the SIS model, $p_J$ is uniform. Solid lines are
      solutions of Eq. (\ref{hu1}) using the population dynamics algorithm with $5 \times 10^4$ paths and initial condition $x_i(0) = 10^{-3}$.
      Symbols denote numerical simulations of Eq. (\ref{uds}) for an ensemble of $10$ random
      networks generated from the configuration model with $N=4000$ nodes. Vertical bars are the standard deviation of the macroscopic observables.
}
\label{compare}
\end{center}
\end{figure}

Equation (\ref{hu1}) is formally similar to other
distributional equations appearing in the study of sparse disordered systems \cite{Mezard2001,Mezard2003,Kuhn2008,Metz2019,AbouChacra1973,MezardBook}. Therefore, we
can numerically solve this equation using the population
dynamics algorithm \cite{Mezard2001,Kuhn2008,Metz2019}.
In the standard version of this algorithm \cite{Mezard2001,Metz2019}, a probability density is parametrized by a population of stochastic variables. 
Here, we generalize the algorithm  to calculate $\mathcal{P}[x]$ by introducing a
population of dynamical trajectories. At each iteration step, a single path is chosen randomly from the population
and updated according to the differential equation 
imposed by the Dirac-$\delta_F$ in Eq. (\ref{hu1}).
After sufficient iterations, the population
of paths converges to a stationary distribution, providing a numerical solution for $\mathcal{P}[x]$.
A detailed account of the algorithm is in \cite{Note1}.

In Fig. \ref{compare}, we compare the solutions of Eq. (\ref{hu1}) with numerical simulations of the original coupled dynamics, Eq. (\ref{uds}), for finite $N$. The panels
showcase the time-evolution of the mean $m(t)$ and the standard deviation $\sqrt{q(t) - m^2(t)}$ of the microscopic variables for different network
topologies and three  models of table \ref{TableModels}: the NN model of \cite{Sompolinsky1988}, the LV model, and the SIS model of epidemic
spreading. The results in Fig. \ref{compare} are for Poisson and geometric indegrees, where  $p_{{\rm in},k}$ is given by
\begin{equation}
  p_{{\rm in},k} = \frac{c^k e^{-c}}{k!} \quad {\rm and} \quad p_{{\rm in},k} = \frac{c^k}{(c+1)^{k+1}},
  \label{degru}
\end{equation} 
respectively. In \cite{Note1}, we compare the solutions of Eq. (\ref{hu1}) with numerical simulations for two additional cases: the NN model
on networks with power-law indegree distributions and the SIS model at the epidemic threshold.
In all cases, the agreement between our theoretical results for $N \rightarrow \infty$ and numerical simulations for large $N$ is
excellent, confirming the exactness of Eq. (\ref{hu1}).

An important question is whether Eq. (\ref{hu1}) recovers the analytic results
of fully-connected models as $c \rightarrow \infty$ \cite{Sompolinsky1988,Crisanti2018,Galla2018,Mastrogiuseppe2018}.
By taking the limit $c \rightarrow \infty$ {\it after} the thermodynamic limit $N \rightarrow \infty$, we are
  effectively considering the scaling regime where $c \propto N^a$, with $0 < a < 1$ \cite{Metz2020,Metz2022,Silva2022,Ferreira2023}.
We show in \cite{Note1} that, in the limit $c \rightarrow \infty$, degree fluctuations remain significant, and $\mathcal{P}[x]$ depends on the
full distribution $\nu_{\rm in}(\kappa)$ of rescaled indegrees $\kappa_i=K_i/c$ ($i=1,\dots,N$). The well-known effective dynamics
on fully-connected networks is only recovered when $\nu_{\rm in}(\kappa) = \delta(\kappa-1)$.
This breakdown in the universality of fully-connected models 
due to degree fluctuations was anticipated in \cite{Metz2020,Metz2022,Silva2022,Ferreira2023}.


\paragraph{Phase diagram of neural networks.}

To demonstrate the strength of Eq. (\ref{hu1}) and its concrete applications, we derive the phase diagram of
the NN model on sparse directed networks. In this context, $x_i(t) \in \mathbb{R}$ represents the
synaptic current at neuron $i$, $f(x) = x$, and $g(x,x^{\prime})=\tanh{\left( x^{\prime} \right)}$.
Let us order the complex eigenvalues $\{ \lambda_\alpha \}_{\alpha=1,\dots,N}$ of $\boldsymbol{A}$ according
to their real parts as ${\rm Re} \lambda_1 \geq {\rm Re} \lambda_2 \geq \dots \geq {\rm Re} \lambda_N$.
A linear stability analysis of Eq. (\ref{uds}) shows that $\boldsymbol{x}=0$ is stable if ${\rm Re} \lambda_{1} < 1$.
Based on analytic results for the spectra of directed networks for $N \rightarrow \infty$ \cite{Neri2016,Neri2020,Metz2021}, we find
that the trivial fixed-point
is stable provided $c < c_{\rm stab}$, where
\begin{equation}
  c_{\rm stab} =
  \begin{cases}
    1/\mu_J & \,\,\text{if $c >  1 + \sigma_J^2/\mu_J^2  $}, \\
    1/(\sigma_J^2 + \mu_J^2) & \text{ if $c \leq 1 + \sigma_J^2/\mu_J^2 $  }.
  \end{cases}
  \label{stab}
\end{equation}
The two distinct regimes in Eq. (\ref{stab}) result from the gap-gapless transition in the
spectrum of $\boldsymbol{A}$ \cite{Neri2016}.
For $c > 1 + \sigma_J^2/\mu_J^2 $, the spectral gap $|\lambda_{1} - \lambda_{2}|$ remains
finite as $N \rightarrow \infty$, while it vanishes for $c \leq 1 + \sigma_J^2/\mu_J^2 $.
Equation (\ref{stab}) holds for $\mu_J >0$ and $c > 1$. The condition $c>1$ ensures that directed networks with degree
  distribution $p_{k,\ell}= p_{{\rm in},k} p_{{\rm out},\ell}$ contain a giant
  strongly connected component \cite{Timar2017,Kryven2016}, implying the existence of a continuous part in the eigenvalue distribution of $\boldsymbol{A}$ \cite{Neri2020}.
\begin{figure}[H]
  \begin{center}
    \includegraphics[scale=1.2]{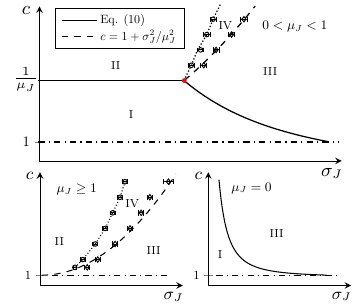}
    \caption{Phase diagrams $(c,\sigma_J)$ of the NN model for $\mu_J=0$, $\mu_J \geq 1$, and $0 < \mu_J < 1$.
      The average synaptic current $m(t)$ relaxes to trivial and
      nontrivial fixed-points in phases I and II, respectively.
      In the chaotic phases III and IV, $m(t)$ exhibits aperiodic oscillations around zero and
      nonzero values, respectively (see Fig. \ref{trans1}). The dash-dotted line
      at $c=1$ marks the percolation threshold, while the dashed curve identifies the gap-gapless transition.
      Circles and diamonds represent numerical results for the transition lines delimiting phase IV (see the main text), obtained
      from solutions of Eq. (\ref{hu1}) using population dynamics with $5 \times 10^{4}$ paths, Poisson indegrees, and two
      values of $\mu_J$ ($\mu_J=1/3$ and $\mu_J = 3/2$). The red dot marks $(c^{*},\sigma^{*}_{J}) = ( \mu_J^{-1}, \sqrt{\mu_J(1- \mu_J)} )$.
}
\label{phasediag}
\end{center}
\end{figure}

We emphasize that the linear stability analysis of the trivial solution
provides no information about the relaxation dynamics or the  stationary solutions that emerge when $\boldsymbol{x}=0$ becomes unstable.
Therefore, we study the dynamics and the stationary states of the NN model by solving
Eq. (\ref{hu1}). Figure \ref{phasediag} presents the resulting phase diagrams for different regimes of $\mu_J$.
In phase I, $m(t)$ relaxes exponentially fast to the trivial solution $m=0$, while in phase II, $m(t)$ evolves to a nonzero fixed-point.
In phases III and IV, the neural network exhibits chaotic activity, characterized by
slow and aperiodic oscillations of $m(t)$ \cite{Sompolinsky1988,Mastrogiuseppe2018}.
The stability lines that delimit phase I are universal, as they depend only on
the first and second moments of $p_J$ and $p_{{\rm in},k}$.

\begin{figure}[H]
  \begin{center}
    \hspace{-0.5cm}
    \includegraphics[scale=1.15]{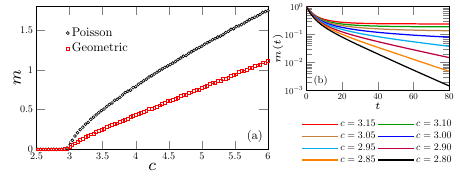}
    \caption{Continuous transition between the fixed-point phases for the NN model on directed random networks. The coupling
      strengths follow a Gaussian distribution with mean $\mu_J = 1/3$ and standard deviation $\sigma_J=0.1$. The
      results are derived from Eq. (\ref{hu1}) using population dynamics with $5 \times 10^4$ paths.
      (a) Fixed-point solution $m(t) = m$ as a function of $c$ for Poisson and geometric indegrees (Eq. (\ref{degru})). (b) Relaxation dynamics of $m(t)$ across
      the transition for Poisson indegrees and initial condition $x_i(0)=1$. For $c < \mu_{J}^{-1}$, $m(t)$ relaxes exponentially to $m=0$ as $t \rightarrow \infty$.
}
\label{trans}
\end{center}
\end{figure}

\begin{figure}
    \includegraphics[scale=1.56]{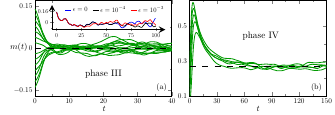}
    \caption{Dynamics of $m(t)$ inside the chaotic phases for directed networks
      with Poisson indegrees and a Gaussian distribution $p_J$ with mean $\mu_J=1/3$.
      The results follow from the solutions of  Eq. (\ref{hu1}) using  population dynamics with $5 \times 10^4$ paths and several initial conditions.
      The dashed lines indicate the value of $M$, Eq. (\ref{M}), characterizing the chaotic attractor in each phase.
      (a) $c=2.5$ and $\sigma_J=2$. Inset: dynamics of $m(t)$ obtained from the {\it deterministic} Eq. (\ref{uds}) for a
      single network instance ($N=4000$), initial condition $x_i(0) = 0.15-\epsilon$, and different $\epsilon$. 
      (b) $c=5$ and $\sigma_J=0.62$.
}
\label{trans1}
\end{figure}

Figure \ref{trans} characterizes the transition between the fixed-point phases I and II. As $c$ approaches the
critical mean degree $c^{*}=\mu_J^{-1}$ from above, the nontrivial fixed-point $m = m(t)$ vanishes continuously. The
critical point $c^{*}$ is independent of $p_{{\rm in}, k}$ and $m(t)$ relaxes exponentially fast inside phases I and II.
Due to critical slowing down, the numerical solution of Eq. (\ref{hu1}) becomes
computationally more demanding near $c^{*}$. In phase II, the neuronal firing rates evolve
to a stationary distribution with a finite variance.
Figure \ref{trans1} shows the dynamics of $m(t)$ within the chaotic phases for
several initial conditions. After a transient time $T_{\rm tr}$, $m(t)$ stabilizes into an
attractor, oscillating around zero in phase III and around a nonzero value in phase IV. The
inset in Fig.  \ref{trans1}(a) demonstrates the sensitivity of $m(t)$ to small perturbations in
the initial conditions, characteristic of deterministic chaos \cite{BakerBook}. By
solving Eq. (\ref{hu1}) and numerically computing the temporal averages \cite{Gelbrecht2020}      
\begin{align}
  M &=  \frac{1}{T-T_{\rm tr}}\int_{T_{\rm tr}}^T dt \, m(t),  \label{M} \\
  \Delta^2 &=  \frac{1}{T-T_{\rm tr}}\int_{T_{\rm tr}}^T dt \, \left[ M - m(t)  \right]^2,
  \label{Delta}
\end{align}
for $T \gg 1$, we estimate the transition lines delimiting phase IV \cite{Note1}.
For $T \rightarrow \infty$, the transition between
phases II and IV is determined by $\Delta$, since $\Delta = 0$ in the fixed-point phases
and $\Delta >0$ in the chaotic phases.
The parameter $M$ distinguishes the chaotic phases: $M=0$ in phase III and $M \neq 0$ in phase IV (see Fig. \ref{trans1}).
Our numerical results for
the transition between phases III and IV are consistent with the dashed line in Fig. \ref{phasediag}, suggesting that
this transition is governed by the gap $|\lambda_1-\lambda_2|$ in the network spectrum. Nevertheless, for $c \rightarrow \infty$ this transition
to the chaotic phase III slightly deviates from the gap-gapless transition \cite{Mastrogiuseppe2018}, particularly for large $\mu_J$.

\paragraph{Conclusions.}

We have developed a dynamical mean-field theory of complex systems on sparse directed networks, deriving an exact path-probability
equation for the effective dynamics in the limit $N \rightarrow \infty$.
Unlike sparse models with bidirected edges \cite{Hatchett2004,Neri2009}, our
path-probability equation can be numerically solved using population dynamics \cite{Metz2019}. We confirmed the exactness of our general solution
by comparing it with numerical simulations of fundamental models in the study of epidemic spreading, neural networks, and ecosystems.
Finally, we applied our solution to determine the complete phase diagram of the sparse and directed
version of the canonical neural network model of \cite{Sompolinsky1988,Mastrogiuseppe2018}.

The numerical solution of Eq. (\ref{hu1}) does not require generating networks from the configuration model \cite{Fosdick2018}, providing 
moderate computational advantages over large-scale simulations of Eq. (\ref{uds}) on networks \cite{Silva2020}.
Beyond its numerical applications, Eq. (\ref{hu1}) provides
a foundational framework for studying the stationary states \cite{Crisanti2018}, computing correlation functions \cite{Marti2018}, deriving
approximate dynamical equations for macroscopic order-parameters, and developing systematic perturbative approaches \cite{Metz2014,Hertz2017}.

Our work paves the way for exploring the role of sparse heterogeneous networks on the dynamics
of ecosystems, coupled oscillators, epidemic spreading, and beyond. 
Future works include determining the phase diagram of the
sparse Lotka-Volterra model \cite{Bunin2017}, the influence of external
noise on phase diagrams \cite{Schuecker2018},
and the role of network heterogeneities on the critical exponents of complex systems \cite{Ferreira2023,Pirey2024}.
Lastly, it would
be interesting to connect the present formalism with the cavity method in \cite{Tarabolo2024}.

\acknowledgements

F. L. M. thanks Tuan Minh Pham and Alessandro Ingrosso for many stimulating discussions. The author also thanks Thomas Peron
and Tuan Minh Pham for their valuable comments on the manuscript. F. L. M. acknowledges support from CNPq (Grant No. 402487/2023-0) and the ICTP
  through the Associates Programme (2023-2028).

\bibliography{bibliography}

 \end{document}


\date{\today}
\title{\bf Supplemental material for ``Dynamical Mean-Field Theory of Complex Systems on Sparse Directed Networks''}
\author{Fernando L. Metz}
 \affiliation{Physics Institute, Federal University of Rio Grande do Sul, 91501-970 Porto Alegre, Brazil}

\maketitle


\section{Introduction}  
        
In this supplemental material, we show how to apply dynamical mean-field theory (DMFT) to solve nonlinear systems of coupled differential equations governing a broad
range of dynamical processes on sparse directed graphs. The microscopic variables $x_1(t),\dots,x_N(t)$ evolve in time according to the coupled differential equations
\begin{equation}
  \dot{x}_i(t) = - f(x_i) + \sum_{j=1}^N C_{ij}J_{ij} g(x_i,x_j) + h_i(t) +  \xi_i(t),
  \label{uds}
\end{equation}
with $i=1,\dots,N$. The variable $\xi_i(t)$ represents an uncorrelated Gaussian noise with mean zero and variance $\sigma^2$ (in the
main text, we present results for $\xi_i(t) = 0$).
Unlike the model definitions in the main text, here we have included, for technical purposes, an external field $h_i(t)$ in the dynamical equations, which will be
set to zero at the end of the calculation. The functions $f(x)$ and $g(x,x^{\prime})$ specify the model under study. Table I of the main text
shows five paradigmatic examples of complex systems modeled by Eq. (\ref{uds}).

The binary variable $C_{ij} \in \{ 0 , 1\}$ tells us whether there is a directed edge
pointing from node $j$ to node $i$ in the network, while $J_{ij}$ represents the strength of the directed interaction $j \rightarrow i$.
We set $C_{ii}=0$. The off-diagonal entries $\{ C_{ij} \}_{i \neq j}$ encode the network topology, and they are drawn  from the joint probability distribution
\begin{equation}
  \mathbb{P}(\{ C_{ij} \}) = \frac{1}{\mathcal{N}}  \prod_{i \neq j=1}^N \left[ \frac{c}{N} \delta_{C_{ij},1} + \left(1 - \frac{c}{N} \right) \delta_{C_{ij},0} \right]
  \prod_{i=1}^N \delta_{K_i, \sum_{j=1}^N C_{ij} } \delta_{L_i, \sum_{j=1}^N C_{ji} },
  \label{gesd}
\end{equation}  
where $\mathcal{N}$ is the normalization factor. The local variables $K_i= \sum_{j=1}^N C_{ij}$ and $L_i = \sum_{j=1}^N C_{ji} $ denote, respectively, the indegree
and the outdegree of node $i$. The joint probability distribution of indegrees $k$ and outdegrees $\ell$  is formally defined as
\begin{equation}
p_{k,\ell} =  \lim_{N \rightarrow \infty} \frac{1}{N} \sum_{i=1}^{N} \delta_{K_i,k} \delta_{L_i,\ell},
\end{equation}  
while the mean degree $c$ is given by
\begin{equation}
c = \sum_{k,\ell=0}^\infty k \, p_{k,\ell}   = \sum_{k,\ell=0}^\infty \ell \, p_{k,\ell} .
\end{equation}  
The parameter $c$ controls the average number of links that enter or leave a node. 
The distribution of Eq. (\ref{gesd}) defines an ensemble of directed random networks in which a directed link $j \rightarrow i$ is randomly placed between
nodes $i$ and $j$ with probability $c/N$, subject to the local constraints imposed by the prescribed degree sequences $K_1,\dots,K_N$ and $L_1,\dots,L_N$.

The degrees
at different nodes are independent and identically distributed random variables drawn from the joint distribution $p_{k,\ell}$.
However, the pair of degrees $(K_i,L_i)$ at node $i$ might be correlated. If $K_i$ and $L_i$ are statistically independent, then the joint degree
distribution $p_{k,\ell}$ factorizes as $p_{k,\ell} = p_{{\rm in},k} p_{{\rm out},\ell}$. Here we solve the dynamics for an arbitrary $p_{k,\ell}$, while in the main
text we present the final solution for the factorized case $p_{k,\ell} = p_{{\rm in},k} p_{{\rm out},\ell}$.
The coupling
strengths $\{ J_{ij} \}$ are independent and identically distributed random variables drawn from an arbitrary distribution $p_J$ with mean $\mu_J$ and variance $\sigma_J^2$. Thus, the distributions $p_{k,\ell}$
and $p_J$ fully specify the ensemble of networks.

In the next section, we detail how to solve the current model, in the limit $N \rightarrow \infty$, using dynamical
mean-field theory (DMFT). We emphasize that we solve the model for {\it finite} $c$, which characterizes a genuine sparse interacting system, where the average number
of edges per node remains finite in the thermodynamic limit.
As we show below, the main outcome of DMFT for finite $c$ is a self-consistent equation for the path-probability $\mathcal{P}[x]$ encoding
the effective dynamics of a single dynamical variable. The solution of this equation not only allows to compute macroscopic
observables, but it also enables to follow the time-evolution of the full probability distribution of the microscopic degrees of freedom.
In section \ref{popdyn}, we explain how to generalize the population dynamics algorithm  \cite{Mezard2001,Mezard2003,Metz2019}  to solve the self-consistent
equation for the path-probability $\mathcal{P}[x]$. We also present a detailed account of this algorithm, which we hope will be useful
for future works, and we explain how to estimate the transition lines in Fig. 2 of the main text. Finally, in
section \ref{high}, we calculate the high-connectivity limit $c \rightarrow \infty$ of $\mathcal{P}[x]$ and
discuss its convergence towards fully-connected models. In particular, we show that the expression for $\mathcal{P}[x]$ of fully-connected models
is only recovered if the variance of rescaled degrees goes to zero. In other words, the dynamics of fully-connected models
is not universal with respect to degree fluctuations.


\section{Dynamical mean-field theory for models on sparse directed networks}
\label{calculation}

In this section we explain how to solve the dynamics of the model in the limit $N \rightarrow \infty$ using dynamical mean-field theory. Let $\boldsymbol{x}(t) = (x_1(t),\dots,x_N(t))$ denotes
the global state of the system at time $t$, and let $\mathcal{P}[\boldsymbol{x}]$ be the functional probability density of observing a continuous trajectory of the
global state $\boldsymbol{x}(t)$ in a finite time interval. By defining the distribution $p_{0}(\boldsymbol{x}(0))$ of the initial state $\boldsymbol{x}(0)$, the
path-probability $\mathcal{P}[\boldsymbol{x}]$ can be formally written as
\begin{equation}
  \mathcal{P}[\boldsymbol{x}] = p_{0}(\boldsymbol{x}(0))  \int \left( \prod_{i=1}^N D \xi_i  \right) \mathcal{P}_{\sigma} [\boldsymbol{\xi}] \prod_{i=1}^N
  \delta_F \Bigg[   \dot{x}_i + f(x_i) - \sum_{j=1}^N C_{ij}J_{ij} g(x_i,x_j) - h_i(t) +  \xi_i(t)  \Bigg],
  \label{trw}
\end{equation}
where $\delta_F$ is a Dirac-$\delta$ functional, and the generic symbol $D y$ represents a functional integration measure over all possible functions $y(t)$ in
the prescribed time domain.
The object $\mathcal{P}_{\sigma} [\boldsymbol{\xi}]$ is the path-probability of the Gaussian noise
\begin{equation}
\mathcal{P}_{\sigma} [\boldsymbol{\xi}] = \frac{1}{\mathcal{N}_{\sigma}} \exp{\left( - \frac{1}{2 \sigma^2} \int dt \sum_{i=1}^N \left( \xi_i(t) \right)^2   \right)},
\end{equation}  
with $\mathcal{N}_{\sigma}$ the normalization constant. The integrals over time run over a fixed time interval, but we omit the
limits of integration, here and elsewhere, for the sake of simplifying the notation.
The starting point of DMFT is the introduction of the generating functional $\mathcal{Z} \left[ \boldsymbol{\psi},\boldsymbol{h} \right]$
of the path-probability $\mathcal{P}[\boldsymbol{x}]$, 
\begin{equation}
  \mathcal{Z} \left[ \boldsymbol{\psi}, \boldsymbol{h}  \right] = \int \left( \prod\limits_{i=1}^N D x_i   \right) \mathcal{P}[\boldsymbol{x}] \, e^{i \int dt \sum\limits_{i=1}^N x_i(t) \psi_i(t)  },
\end{equation}
where $\boldsymbol{\psi}(t) = (\psi_1(t),\dots,\psi_N(t))$ and $\boldsymbol{h}(t) = (h_1(t),\dots,h_N(t))$.
The derivatives of $\mathcal{Z} \left[ \boldsymbol{\psi}, \boldsymbol{h} \right]$ with respect to the
sources $\psi_1(t),\dots,\psi_N(t)$ yield the average of products of $x_1(t),\dots,x_N(t)$ over the dynamical process
specified by Eq. (\ref{uds}). For instance, the moments of $x_i(t)$ in the absence of external fields ($h_i(t)=0$) follow from
\begin{equation}
  \langle x^{n}_{i}(t) \rangle = (- i)^n \lim_{\boldsymbol{\psi} \rightarrow 0} \lim_{\boldsymbol{h} \rightarrow 0} \frac{\delta^{n} \mathcal{Z} \left[ \boldsymbol{\psi},\boldsymbol{h} \right]  }{ \delta \psi_{i}^{n}(t)}.
  \label{yuyu4}
\end{equation}  
In particular, the first two moments yield the time-evolution of the following macroscopic observables
\begin{eqnarray}
  m(t) = \lim_{N \rightarrow \infty} \frac{1}{N} \sum_{i=1}^N \langle x_{i}(t) \rangle, \label{gugu1} \\
  q(t) = \lim_{N \rightarrow \infty} \frac{1}{N} \sum_{i=1}^N \langle x^{2}_{i}(t) \rangle. \label{gugu2}
\end{eqnarray} 
By construction, the generating functional fulfills the normalization condition $\mathcal{Z} \left[0,\boldsymbol{h} \right] =1$.

By inserting Eq. (\ref{trw}) into the definition of $\mathcal{Z} \left[ \boldsymbol{\psi}, \boldsymbol{h}  \right]$, using a functional integral representation
of $\delta_F$, and then integrating over the Gaussian noise $\boldsymbol{\xi}(t)$, we obtain the following expression
\begin{eqnarray}
   \mathcal{Z} \left[ \boldsymbol{\psi},\boldsymbol{h} \right] &=& \int \left( \prod_{i=1}^N D x_i D \hat{x}_i \right)  p_{0}(\boldsymbol{x}(0))
   \exp{ \Bigg( - i \sum\limits_{i \neq j=1}^N C_{ij} J_{ij} \int dt  \, \hat{x}_i (t) \, g \left( x_i(t),x_j(t) \right) \Bigg)} \nonumber \\
   &\times& \exp{\Bigg( i \int dt \sum_{i=1}^N S_i \left( x_i (t), \hat{x}_i(t),  \psi_i(t), h_i(t)  \right)   \Bigg) },
  \label{huji}
\end{eqnarray}  
where the single-site action $S_i$ is defined as
\begin{equation}
  S_i \left[ x_i (t), \hat{x}_i(t), \psi_i(t) , h_i(t) \right] = x_i(t) \psi_i(t) + \frac{i \sigma}{2} \hat{x}_i^2(t) + \hat{x}_i(t) \left[ \dot{x_i}(t) + f \left( x_i(t) \right) - h_i (t)   \right].
  \label{opop1}
\end{equation}  
%
According to Eq. (\ref{huji}), the external
sources $h_i(t)$ and $\psi_i(t)$ are coupled, respectively, to the dynamical variables $\hat{x}_i(t)$ and $x_i(t)$ at
node $i$. Both fields $h_i(t)$ and $\psi_i(t)$ are important to simplify the saddle-point equations and identify the
physical meaning of the order-parameters.


\subsection{The saddle-point integral}

The aim of DMFT is to compute, in the limit $N \rightarrow \infty$, the disorder averaged generating
functional, $\langle \mathcal{Z} \left[ \boldsymbol{\psi},\boldsymbol{h} \right] \rangle_{\{ C_{ij},J_{ij} \}}$. The symbol $\langle \dots \rangle_{\{ C_{ij},J_{ij} \}}$ represents the average
over both the graph structure $\{ C_{ij} \}$ and the coupling strengths $\{ J_{ij} \}$. In this subsection, we explain how to reduce the calculation
of $\langle \mathcal{Z} \left[ \boldsymbol{\psi},\boldsymbol{h} \right] \rangle_{\{ C_{ij},J_{ij} \}}$ to the solution of a saddle-point
integral in the limit $N \rightarrow \infty$.


By taking the average of Eq. (\ref{huji}) over $\{ C_{ij} \}$ and $\{ J_{ij} \}$, we can write $\langle \mathcal{Z} \left[ \boldsymbol{\psi},\boldsymbol{h} \right] \rangle_{\{ C_{ij},J_{ij} \}}$
as follows
\begin{equation}
  \langle \mathcal{Z} \left[ \boldsymbol{\psi},\boldsymbol{h} \right] \rangle_{\{ C_{ij},J_{ij} \}} = \int \left( \prod_{i=1}^N D x_i D \hat{x}_i \right)
  p_{0}(\boldsymbol{x}(0))
 \mathcal{F}[ \boldsymbol{x}, \boldsymbol{\hat{x}}]  \exp{\left( i \int dt \sum_{i=1}^N S_i \left( x_i (t), \hat{x}_i(t), \psi_i(t) , h_i(t) \right)   \right)},
 \label{tqc}
\end{equation}  
where
\begin{equation}
  \mathcal{F}[ \boldsymbol{x}, \boldsymbol{\hat{x}}] =   \left\langle  \prod \limits_{i \neq j=1}^N
  \exp{\left( - i  C_{ij} J_{ij} \int dt \, \hat{x}_i (t) g \left( x_i(t),x_j(t) \right) \right)} \right\rangle_{\{ C_{ij},J_{ij} \}}.
  \label{gppx}
\end{equation}  
By using the integral representation of the Kronecker-$\delta$
\begin{equation}
\delta_{n,m} = \int_{0}^{2 \pi} \frac{du}{2 \pi} \exp{\left[ i u (n-m) \right]} \quad (n,m \in \mathbb{N}),
\end{equation}  
we can factorize the joint distribution $\mathbb{P}(\{ C_{ij} \})$, Eq. (\ref{gesd}), in terms of  a product over pairs of nodes
\begin{equation}
  \mathbb{P}(\{ C_{ij} \}) = \frac{1}{\mathcal{N}} \int\limits_{0}^{2 \pi} \left( \frac{d u_i d v_i}{4 \pi^2} \right) e^{i \sum\limits_{i=1}^N  \left( K_i u_i + L_i v_i  \right)  }  
  \prod_{i \neq j =1}^N \left[ \frac{c}{N} \delta_{C_{ij},1} + \left(  1 - \frac{c}{N} \right)  \delta_{C_{ij},0} \right] e^{- C_{ij} (u_i + v_j)}.
  \label{gwhh1}
\end{equation}  
This expression for $ \mathbb{P}(\{ C_{ij} \})$ allows us to calculate the average over $\{ C_{ij} \}$ in Eq. (\ref{gppx}), leading to the result
\begin{equation}
  \mathcal{F}[ \boldsymbol{x}, \boldsymbol{\hat{x}}] \simeq \int_{0}^{2 \pi} \left( \prod_{i=1}^N \frac{d u_i  d v_i}{ 4 \pi^2}  \right) 
  \exp{\left( i \sum \limits_{i=1}^N \left( u_i K_i + v_i L_i  \right) +
     \frac{c}{N} \sum_{ij=1}^N e^{-i \left( u_i + v_j \right)}  \left\langle  e^{- i J \int dt \, \hat{x}_i(t) \,   g \left( x_i(t), x_j(t) \right)    }  \right\rangle_J  \right)   },
  \label{pqnf}
\end{equation}  
valid for large $N$. We have retained only terms of $\mathcal{O}(N)$ in the exponent of Eq. (\ref{pqnf}), since these are precisely the terms
that will contribute to the solution of the saddle-point integral for $N \rightarrow \infty$. In addition, irrelevant terms that arise from
the normalization factor in Eq. (\ref{gesd}) and which are independent of $\{ \hat{x}_i(t) , x_i(t) \}$ have been
neglected when writing Eq. (\ref{pqnf}). The symbol $\langle \dots \rangle_J$ denotes the
average over the coupling strength $J$ between a single pair of nodes.

In contrast to fully-connected Gaussian models, where the disorder average of
$\mathcal{Z} \left[ \boldsymbol{\psi},\boldsymbol{h} \right]$ yields a quadratic functional of the dynamical variables \cite{Sompolinsky1988,Crisanti2018}, here the exponent in Eq. (\ref{pqnf}) contains
an exponential functional of $\{ \hat{x}_i(t) , x_i(t) \}$. This non-quadratic behaviour renders the problem
technically more challenging in comparison to fully-connected models. In spite of that, we can still decouple sites by introducing the functional order-parameters
\begin{align}
&P[x] = \frac{1}{N} \sum_{i=1}^N \delta_F \left[ x(t) - x_i(t)  \right] e^{-i v_i}, \label{o1} \\
&W[x,\hat{x}] = \frac{1}{N} \sum_{i=1}^N \delta_F \left[ x(t) - x_i(t)  \right]  \delta_F \left[ \hat{x}(t) - \hat{x}_i(t)  \right]   e^{-i u_i}.  \label{o2}
\end{align} 
%
Indeed, substituting Eq. (\ref{pqnf}) in Eq. (\ref{tqc}), we can
write $\langle \mathcal{Z} \left[ \boldsymbol{\psi},\boldsymbol{h} \right] \rangle_{\{ C_{ij},J_{ij} \}}$ as a functional integral
over $P[x]$ and $W[x,\hat{x}]$, namely
\begin{align}
&\langle \mathcal{Z} \left[ \boldsymbol{\psi},\boldsymbol{h} \right] \rangle_{\{ C_{ij},J_{ij} \}} \simeq 
\int \left( \prod_{i=1}^N D x_i D \hat{x}_i \right) p_{0}(\boldsymbol{x}(0)) \int_{0}^{2 \pi} \left( \prod_{i=1}^N \frac{d u_i  d v_i}{ 4 \pi^2}  \right)
e^{ i \int dt \sum\limits_{i=1}^N S_i \left( x_i (t), \hat{x}_i(t), \psi_i(t), h_i(t) \right)  + i \sum \limits_{i=1}^N \left( u_i K_i + v_i L_i  \right) } \nonumber \\
& \times \int D P \, D W \, \delta_F \left[  P[x] -  \frac{1}{N} \sum_{i=1}^N \delta_F \left[ x(t) - x_i(t)  \right] e^{-i v_i} \right]
\delta_F \left[  W[x,\hat{x}] -      \frac{1}{N} \sum_{i=1}^N \delta_F \left[ x(t) - x_i(t)  \right]  \delta_F \left[ \hat{x}(t) - \hat{x}_i(t)  \right]   e^{-i u_i} \right] \nonumber \\
& \times \exp{\left( c N \int Dx Dx^{\prime} D \hat{x} \,  W[x,\hat{x}] P[x^{\prime}]   \left\langle  e^{- i J \int dt \,  \hat{x}(t) \,   g \left( x(t), x^{\prime}(t) \right)    }  \right\rangle_J    \right)}.
\label{fdsfc}
\end{align}  
By assuming that the distribution of the initial state factorizes as $p_{0}(\boldsymbol{x}(0)) = \prod_{i=1}^N p_0(x_i(0))$, and
introducing the conjugate order-parameters $\hat{P}[x]$ and $\hat{W}[x,\hat{x}]$ by means of the functional Fourier transforms,
\begin{align}
 & \delta_F \left[  P[x] -  \frac{1}{N} \sum_{i=1}^N \delta_F \left[ x(t) - x_i(t)  \right] e^{-i v_i} \right] = \int D \hat{P} \,
  e^{i \int D x \hat{P}[x] P[x]  - \frac{i}{N} \sum\limits_{i=1}^N \hat{P}[x_i] e^{-i v_i}   }, \nonumber \\
 & \delta_F \left[  W[x,\hat{x}] -      \frac{1}{N} \sum_{i=1}^N \delta_F \left[ x(t) - x_i(t)  \right]  \delta_F \left[ \hat{x}(t) - \hat{x}_i(t)  \right]   e^{-i u_i} \right] =
  \int D \hat{W} \,
  e^{i \int D x D \hat{x} \hat{W}[x,\hat{x}] W[x,\hat{x}]  - \frac{i}{N} \sum\limits_{i=1}^N \hat{W}[x_i,\hat{x}_i] e^{-i u_i}   }, \nonumber  
\end{align}  
we are able to factorize the exponent of Eq. (\ref{fdsfc}) as a sum of single-site functionals, yielding
\begin{align}
  \langle \mathcal{Z} &\left[ \boldsymbol{\psi},\boldsymbol{h} \right] \rangle_{\{ C_{ij},J_{ij} \}}  \simeq
  \int DP D \hat{P} D W D \hat{W}  \exp{\left( i \int D x \, \hat{P}[x] P[x] + i \int D x \, D \hat{x} \, \hat{W}[x,\hat{x}] W[x,\hat{x}] \right)} \\
  &\times  \exp{\left( c N \int Dx Dx^{\prime} D \hat{x} \,  W[x,\hat{x}] P[x^{\prime}]   \left\langle  e^{- i J \int dt \, \hat{x}(t)    g \left( x(t), x^{\prime}(t) \right)    }  \right\rangle_J    \right)} \nonumber \\
  &\times \exp{\left[ \sum\limits_{i=1}^N \ln{\left(  \int D x D \hat{x} \,\, p_0(x(0)) \int\limits_{0}^{2 \pi}  \frac{d u  d v}{ 4 \pi^2}
        e^{ i \int dt S_i \left( x (t), \hat{x}(t), \psi_i(t), h_i(t) \right)  + i  \left( u K_i + v L_i  \right)  - \frac{i}{N} e^{- i v} \hat{P}[x]  - \frac{i}{N} e^{- i u} \hat{W}[x,\hat{x}]  }    \right) }  \right]}.
  \label{ghgh1}
\end{align}

Finally, we use the power-series representations
\begin{equation}
\exp{\left( - \frac{i}{N} e^{- i v} \hat{P}[x]    \right)} = \sum_{r=0}^\infty \frac{1}{r!} \left( - \frac{i}{N}  \right)^r \left( \hat{P}[x] \right)^r e^{- i v r} 
\end{equation}  
and
\begin{equation}
\exp{\left( - \frac{i}{N} e^{- i u} \hat{W}[x,\hat{x}]    \right)} = \sum_{r=0}^\infty \frac{1}{r!} \left( - \frac{i}{N}  \right)^r \left( \hat{W}[x,\hat{x}] \right)^r e^{- i u r}, 
\end{equation}  
%
which enables to integrate over $u$ and $v$ in Eq. (\ref{ghgh1}). By rescaling the conjugate
order-parameters in the resulting expression as $\hat{P}[x] \rightarrow N \hat{P}[x]$ and $\hat{W}[x,\hat{x}] \rightarrow N \hat{W}[x,\hat{x}]$, we can
express $\left\langle \mathcal{Z} \left[ \boldsymbol{\psi},\boldsymbol{h} \right]\right\rangle_{\{ C_{ij},J_{ij} \}} $ as an integral
over the order-parameters
\begin{equation}
  \left\langle \mathcal{Z} \left[ \boldsymbol{\psi},\boldsymbol{h} \right] \right\rangle_{\{ C_{ij},J_{ij} \}} \simeq \int D P D\hat{P} D W D \hat{W}
  e^{N \Phi [P,\hat{P},W,\hat{W} ] },
  \label{ssppqq}
\end{equation}
with
\begin{align}
  \Phi [P,\hat{P},W,\hat{W} ] &= i \int D x \hat{P}[x] P[x] + i \int D x D \hat{x} \hat{W}[x,\hat{x}] W[x,\hat{x}]
  + c \int D x D x^{\prime} D \hat{x} W[x ,\hat{x}] P[x^{\prime}] \left\langle   e^{- i J \int dt \, \hat{x} (t) g \left( x(t) , x^{\prime}(t) \right)   }   \right\rangle_J \nonumber \\
  &+ \frac{1}{N}  \sum\limits_{i=1}^N \ln \left[ \int D x  D \hat{x} \, p_0(x(0)) \, e^{i \int dt S_i \left( x(t), \hat{x}(t), \psi_i(t), h_i(t)   \right)    }
  \left(- i \hat{P}[x] \right)^{L_i} \left(- i \hat{W}[x,\hat{x}] \right)^{K_i}
    \right].
\end{align}  
The above expression for $\Phi$ omits an additive constant that is independent of the order-parameters. Now
we can find the asymptotic behaviour of the integral in Eq. (\ref{ssppqq}) using the saddle-point method. In the limit $N \rightarrow \infty$, the integral is dominated
by the stationary points of $\Phi [P,\hat{P},W,\hat{W} ]$,
\begin{equation}
  \left\langle \mathcal{Z} \left[ \boldsymbol{\psi},\boldsymbol{h} \right] \right\rangle_{\{ C_{ij},J_{ij} \}} \simeq   e^{N \Phi [P_s,\hat{P}_s,W_s,\hat{W}_s ] },
  \label{juju22}
\end{equation}  
where the functional order-parameters that extremize $\Phi [P,\hat{P},W,\hat{W} ]$ fulfill the saddle-point equations
\begin{align}
  &P_s[x] = \frac{1}{N} \sum\limits_{i=1}^N \frac{L_i}{B_i[h_i,\psi_i]}  \, p_0(x(0)) \left(- i \hat{P}_s[x] \right)^{L_i - 1} \int D \hat{x} \left(- i \hat{W}_s[x,\hat{x}] \right)^{K_i}
  e^{i \int dt \, S_i \left( x(t), \hat{x}(t), \psi_i(t), h_i(t)   \right)    }, \label{pp1} \\
  &W_s[x,\hat{x}] = \frac{1}{N} \sum\limits_{i=1}^N \frac{K_i}{B_i[h_i,\psi_i]} \,  p_0(x(0)) \left(- i \hat{W}_s[x,\hat{x}] \right)^{K_i - 1} \left(- i \hat{P}_s[x] \right)^{L_i}
  e^{i \int dt \, S_i \left( x(t), \hat{x}(t), \psi_i(t), h_i(t)   \right)    }, \label{pp2} \\
  & \hat{P}_s[x] = i c \int D x^{\prime} D \hat{x} \, W_s[x^{\prime}, \hat{x}] \left\langle   e^{- i J \int dt \, \hat{x} (t) g \left( x^{\prime}(t) , x(t) \right)   }   \right\rangle_J, \label{pp3} \\
   & \hat{W}_s[x,\hat{x}] = i c \int D x^{\prime}  \, P_s[x^{\prime}] \left\langle   e^{- i J \int dt \, \hat{x} (t) g \left( x (t) , x^{\prime}(t) \right)   }   \right\rangle_J, \label{pp4}
\end{align}  
with
\begin{equation}
B_i[h_i,\psi_i] = \int D x D \hat{x} \, p_0(x(0)) \,  e^{i \int dt \, S_i \left( x(t), \hat{x}(t), \psi_i(t), h_i(t)   \right)    } \left(- i \hat{P}_s[x] \right)^{L_i} \left(- i \hat{W}_s[x,\hat{x}] \right)^{K_i}.
\end{equation}  
The self-consistent Eqs. (\ref{pp1}-\ref{pp4}) are derived from the application of the stationarity conditions
\begin{equation}
  \frac{\delta \Phi}{\delta P[x]} \bigg{|}_s = \frac{\delta \Phi}{\delta \hat{P}[x]} \bigg{|}_s
  = \frac{\delta \Phi}{\delta W[x,\hat{x}]} \bigg{|}_s = \frac{\delta \Phi}{\delta \hat{W}[x,\hat{x}]} \bigg{|}_s = 0, 
\end{equation} 
where the notation $(\dots)\big{|}_s$ means that $P[x]=P_s[x]$, $\hat{P}[x]=\hat{P}_s[x]$, $W[x,\hat{x}] = W_s[x,\hat{x}]$ and $\hat{W}[x,\hat{x}] = \hat{W}_s[x,\hat{x}]$.


\subsection{Physical interpretation of the order-parameters}

In this subsection we explain how to distill the physical interpretation of the functional order-parameters, which plays an important role in simplifying the saddle-point equations. 
In addition, we show how Eqs. (\ref{pp1}-\ref{pp4}) can be reduced to a pair of self-consistent equations for the path-probability
that determines the effective dynamics of a single degree of freedom.

First, we set $\psi_i(t)=h_i(t)=0$ in the single-site action $S_i \left[ x(t), \hat{x}(t), \psi_i(t), h_i(t)   \right]$ (see Eq. (\ref{opop1})) and define the following
normalized functional 
\begin{equation}
  \gamma[x,\hat{x}|k,l] =  \frac{ p_0(x(0)) \,  e^{i \int dt \, S\left( x(t), \hat{x}(t)  \right)    } \left(- i \hat{P}_s[x] \right)^{\ell} \left(- i \hat{W}_s[x,\hat{x}] \right)^{k}  }
        { \int D x D \hat{x} \, p_0(x(0)) \, e^{i \int dt \, S\left( x(t), \hat{x}(t)  \right)    } \left(- i \hat{P}_s[x] \right)^{\ell} \left(- i \hat{W}_s[x,\hat{x}] \right)^{k}   }
        \label{juio}
\end{equation}  
for fixed indegree $k$ and outdegree $\ell$.
Our first task is to find out what is the physical interpretation of the above object. With that
in mind, we take the derivative of Eq. (\ref{juju22}) with respect to $\psi_i(t)$ and then we set $\psi_i(t)=h_i(t)=0$, namely
\begin{equation}
 \lim_{h \rightarrow 0} \lim_{\psi \rightarrow 0}  \frac{ \delta \left\langle \mathcal{Z} \left[ \boldsymbol{\psi},\boldsymbol{h} \right] \right\rangle_{\{ C_{ij},J_{ij} \}} }{\delta \psi_i(t)} =
  i \int D x D \hat{x} \, x(t)  \gamma[x,\hat{x}|K_i,L_i].
\end{equation}  
Comparing the above expression with the first moment $\langle x_i(t) \rangle$ defined in Eq. (\ref{yuyu4}), we conclude that
\begin{equation}
  \mathcal{P}[x|k,\ell] = \int D \hat{x}  \, \gamma[x,\hat{x}|k,\ell]
  \label{iu11}
\end{equation}  
is the path-probability describing the effective dynamics of a single variable $x(t)$ conditioned to the indegree $k$ and outdegree $\ell$. Clearly, the
path-probability $\mathcal{P}[x]$ of observing a dynamical trajectory of $x(t)$ is obtained by averaging $\mathcal{P}[x|k,\ell]$ over the joint degree distribution
%
\begin{equation}
  \mathcal{P}[x] = \sum_{k,l=0}^{\infty} p_{k,\ell} \mathcal{P}[x|k,\ell].
  \label{dfsf1}
\end{equation}  
%
%
%
Equation (\ref{dfsf1}) defines the central object of dynamical mean-field theory, as $\mathcal{P}[x]$ determines the effective dynamics
of a single degree of freedom in the thermodynamic limit $N \rightarrow \infty$. 

Let us derive a self-consistent equation for $\mathcal{P}[x]$.
The first step consists in relating the functional probability densities, $\mathcal{P}[x|k,l]$ and $\gamma[x,\hat{x}|k,l]$, to the functional order-parameters using the saddle-point equations
of the previous section.
By setting $\psi_i(t)=h_i(t) = 0$ in Eqs. (\ref{pp1}) and (\ref{pp3}), we obtain the following relation
\begin{equation}
  - i P_s[x] \hat{P}_s [x] = \sum\limits_{k,\ell=0}^{\infty} \ell \, p_{k,\ell} \mathcal{P}[x|k,\ell].
  \label{hh11}
\end{equation}  
%
Similarly, combining Eqs. (\ref{pp2}) and (\ref{pp4}) for $\psi_i(t)=h_i(t) = 0$, we find
\begin{equation}
  - i W_s[x,\hat{x}] \hat{W}_s [x,\hat{x}] = \sum\limits_{k,\ell=0}^{\infty} k \, p_{k,\ell} \, \gamma[x,\hat{x}|k,l].
  \label{hh22}
\end{equation}  
Let us now see how the saddle-point equations simplify in the case of directed networks.
By introducing the Fourier transform of $W_s[x,\hat{x}]$,
\begin{equation}
R_s[x|\theta] = \int D \hat{x} \, e^{-i \int dt \hat{x}(t) \theta(t)  } W_s[x,\hat{x}],
\end{equation}  
we rewrite Eq. (\ref{pp3}) as follows
\begin{equation}
  \hat{P}_s[x] = i c \int D x^{\prime}  \Big\langle  R_s[x^{\prime}| J  g \left( x^{\prime} , x \right)  ]   \Big\rangle_J.
  \label{iudt}
\end{equation}  
%
Using the definition of $W_s[x,\hat{x}]$, Eq. (\ref{o2}), we write $R_s[x|\theta]$ as
\begin{equation}
R_s[x|\theta] = \frac{1}{N} \sum_{i=1}^N \delta_F \left[ x(t) - x_i(t)  \right]    e^{-i u_i  -i \int dt \hat{x}_i(t) \theta(t)    }.
\end{equation}
Building on the DMFT for the dynamics of binary-state
variables on directed random graphs \cite{Hatchett2004,Mimura2009}, we can compute the average of the above expression
over $\{ x_i(t), \hat{x}_i(t), u_i \}_{i=1}^N$, with the weight defined by Eq. (\ref{ghgh1}), and show that
\begin{equation}
  \int D x R_s[x|\theta] = 1
  \label{hutw}
\end{equation}  
for arbitrary $\theta(t)$.
Thus, from Eqs. (\ref{iudt}) and (\ref{hutw}), we conclude that
\begin{equation}
  \hat{P}_s[x] = i c,
  \label{yirs}
\end{equation}  
which allows us to simplify the saddle-point
equations. The above expression for $\hat{P}_s[x]$ only holds for networks without bidirected edges \cite{Hatchett2004,Mimura2009}.
Now we can identify the physical meaning of the order-parameter $P_s[x]$. Substituting Eq. (\ref{yirs}) in Eqs. (\ref{iu11}) and (\ref{hh11}), we obtain
\begin{equation}
  P_s[x]  = \sum\limits_{k,\ell=0}^{\infty} \dfrac{\ell \, p_{k,\ell}}{c} \mathcal{P}[x|k],
  \label{yted}
\end{equation}
where $\mathcal{P}[x|k]$ is given by
\begin{equation}
 \mathcal{P}[x|k] =  \int \left( \prod_{j=1}^k D x_j P_s[x_j]  \right) \int \left( \prod_{j=1}^k d J_j p_J(J_j)   \right)
 \int D \xi \,  \mathcal{P}_{\sigma}[\xi]  \, \delta_F \left[\dot{x}(t) + f(x(t)) - \sum_{j=1}^k J_j g\left( x(t), x_j(t) \right) - \xi(t)   \right].
 \label{tfna}
\end{equation} 
We have absorbed the distribution $p_0(x(0))$ of the initial state in the definition of $\delta_F[\dots]$, which constraints
the effective dynamics of $x(t)$. Note, from Eq. (\ref{yted}), that $P_s[x]$ is the average of the
conditioned path-probability $\mathcal{P}[x|k]$ over the degrees $k$ and $\ell$ with the joint distribution
$\frac{l p_{k,l}}{c}$. Equations (\ref{yted}) and (\ref{tfna}) lead to the self-consistent equation for $P_s[x]$
\begin{equation}
 P_s[x]  = \sum\limits_{k,\ell=0}^{\infty} \dfrac{\ell \, p_{k,\ell}}{c} \int \left( \prod_{j=1}^k D x_j P_s[x_j]  \right) \int \left( \prod_{j=1}^k d J_j p_J(J_j)   \right)
 \int D \xi \,  \mathcal{P}_{\sigma}[\xi]  \, \delta_F \left[\dot{x}(t) + f(x(t)) - \sum_{j=1}^k J_j g\left( x(t), x_j(t) \right) - \xi(t)   \right].
 \label{gtwar1}
\end{equation} 
Once we determine $P_s[x]$ from the solutions of the above equation, the path-probability $\mathcal{P}[x]$ follows from Eqs. (\ref{dfsf1}) and (\ref{tfna}), namely
\begin{equation}
  \mathcal{P}[x] = \sum\limits_{k, \ell=0}^{\infty} p_{k,\ell} \int \left( \prod_{j=1}^k D x_j P_{s}[x_j]  \right) \int \left( \prod_{j=1}^k d J_j p_J(J_j)   \right)
  \int D \xi \,  \mathcal{P}_{\sigma}[\xi]  \, \delta_F \left[\dot{x}(t) + f(x(t)) - \sum_{j=1}^k J_j g\left( x(t), x_j(t) \right) - \xi(t)   \right].
  \label{gtwar2}
\end{equation}
Equations (\ref{gtwar1}) and (\ref{gtwar2}) make up the main analytic result of this work. Interestingly, these equations are
formally similar to those  determining the eigenvector distribution corresponding to the leading eigenvalue of sparse directed networks \cite{Neri2020,Metz2021}.
When the degrees $K_i$ and $L_i$ are statistically
independent  at any node $i$, the joint degree distribution factorizes, $p_{k,l}=p_{{\rm in},k} \, p_{{\rm out},\ell}$, and the solution of the dynamics reduces
to a single self-consistent equation for the path-probability:
\begin{equation}
  \mathcal{P}[x] = \sum\limits_{k=0}^{\infty} p_{{\rm in},k} \int \left( \prod_{j=1}^k D x_j \mathcal{P}[x_j]  \right) \int \left( \prod_{j=1}^k d J_j p_J(J_j)   \right) \int D \xi \, \mathcal{P}_{\sigma}[\xi] \,
  \delta_F \left[\dot{x}(t) + f(x(t)) - \sum_{j=1}^k J_j g\left( x(t), x_j(t) \right) - \xi(t)   \right].
  \label{gtwar4}
\end{equation}  
%


\section{Population dynamics algorithm}
\label{popdyn}

Here we explain how to generalize  the population dynamics algorithm \cite{Mezard2001,Kuhn2008,Metz2019} to efficiently solve the self-consistent Eq. (\ref{gtwar4}) for the path-probability describing
dynamical processes on directed random networks with uncorrelated degrees $(K_i,L_i)$. The algorithm is based on the introduction of
a large population of stochastic trajectories that parametrize the path-probability $\mathcal{P}[x]$. These dynamical paths are then consistently
updated according to Eq. (\ref{gtwar4}).
Informally speaking, the population dynamics algorithm is a Monte-Carlo approach to solve self-consistent distributional equations
by iteration, analogous to finding the roots of a transcendental fixed-point equation by direct iteration. In the standard version
of the algorithm \cite{Mezard2001,Mezard2003,Metz2019}, the probability density under study characterizes the statistics
of a single random variable, while here one has to generalize the algorithm to deal with the probability density of functions of time. Each
stochastic trajectory is constrained to the differential equation imposed by the Dirac functional-$\delta_F$ in Eq. (\ref{gtwar4}). The most important aspect of the algorithm
in the present case is that a single update  refers to an {\it entire} stochastic trajectory, rather than to a single variable representing a particular instant
of time. Apart from this nuance, the core of the algorithm remains conceptually the same as in standard applications.

Below  we present a detailed account of the
population dynamics algorithm  in the absence of Gaussian noise ($\sigma=0$). For concreteness, we detail
all steps of the algorithm on the neural network model of \cite{Sompolinsky1988}, where $f(x)=x$ and $g(x,x^{\prime})= \tanh{\left( x^{\prime}  \right)}$.
In this particular setting, Eq. (\ref{gtwar4}) assumes the form
\begin{equation}
 \mathcal{P}[x] = \sum\limits_{k=0}^{\infty} p_{{\rm in},k} \int \left( \prod_{j=1}^k D x_j \mathcal{P}[x_j]  \right) \int \left( \prod_{j=1}^k d J_j p_J(J_j)   \right) 
  \delta_F \Big[\dot{x}(t) + x(t) - h_k\left[ \{ J_j, x_j(t) \} \right]  \Big],
  \label{gtwar5}
\end{equation}  
where we defined the time-dependent local field
\begin{equation}
  h_k\left[ \{ J_j, x_j(t) \} \right] = \sum_{j=1}^k J_j \tanh{\left( x_j(t) \right)}.
  \label{local}
\end{equation}  
The extension of the population dynamics algorithm to other models and to nonzero Gaussian noise ($\sigma > 0$) is straightforward.

\begin{algorithm}[H]
  \label{gteds}
  \caption{Population dynamics algorithm to solve Eq. (\ref{gtwar5}) and determine the macroscopic observables $m(t)$ and $q(t)$ as a
    function of time (see Eqs. (\ref{gugu1}) and (\ref{gugu2})). For simplicity, we illustrate
  the algorithm in conjunction with the Euler method for solving differential equations.}
\begin{minipage}{.96\columnwidth}
\begin{algorithmic}

  \State \textbf{Inputs}: indegree distribution $p_{{\rm in},k}$, distribution $p_J$ of coupling strengths, population size $N_{\rm pop}$, total number $N_{\rm iter}$ of single-trajectory
  updates, total number $T$ of points in a discretized path, and time increment $\epsilon$ of the Euler method.
\State \textbf{Outputs}: macroscopic observables $m(t)$ and $q(t)$ at time $t$. \\

\State Set the indegree distribution  $p_{{\rm in},k}$ and the distribution $p_J(J)$ of coupling strengths.
\State Set the matrix $x[t][i]$ ($i=1,\dots,N_{\rm pop}$ and $t=0,\dots,T$) that stores the state variables along the $i$th dynamical trajectory.
\State Initialize all matrix elements $x[t][i]$. The entries $\{ x[0][i] \}_{i=1,\dots,N_{\rm pop}}$ fix the initial condition.
\Repeat

\State Select a single element $i$ uniformly at random from the indexes $1,...,N_{\rm pop}$.
\State Draw a non-negative random integer $k$ from the distribution $p_{{\rm in},k}$.
\State Create a set $\partial_k$  with $k$ elements selected uniformly at random from the indexes $1,...,N_{\rm pop}$.
\State Create a set with elements $J_1,\dots,J_k$ sampled from the distribution $p_J(J)$. 
\State Update the  local fields $h[1][i],\dots,h[T][i]$ along the trajectory $i$: $h[t][i] \leftarrow \sum_{j \in \partial_k} J_j \tanh(x[t-1][j])$  \Comment{Eq.~\eqref{local}} 
\State Update the $i$th trajectory $x[1][i],\dots,x[T][i]$ (Euler method): $x[t][i] \leftarrow  x[t-1][i] + \epsilon \left(  - x[t-1][i] + h[t][i]  \right)$

\Until{step number $< N_{\rm iter}$}



\Return $m(t) \leftarrow \frac{1}{N_{\rm pop}} \sum\limits_{i=1}^{N_{\rm pop}} x[t][i]$ and  $q(t) \leftarrow \frac{1}{N_{\rm pop}} \sum\limits_{i=1}^{N_{\rm pop}} (x[t][i])^2$
\end{algorithmic}
\end{minipage}
\end{algorithm}

\begin{figure}[H]
  \begin{center}
    \includegraphics[scale=1.2]{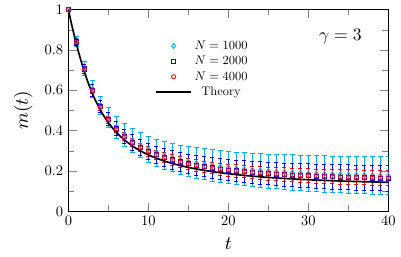}
    \includegraphics[scale=1.2]{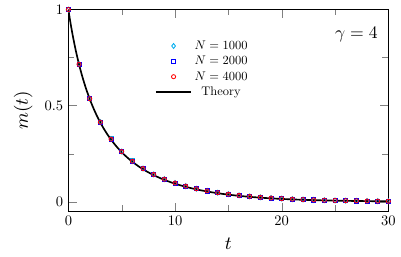}
    \caption{Dynamics of $m(t)$ for the NN model (see table I in the main text) on directed random networks with a power-law
      indegree distribution $p_{{\rm in},k} = \mathcal{N} k^{-\gamma}$, where $\mathcal{N}$ is the normalization factor and $k \geq 2$. The coupling strengths
      are drawn from a Gaussian distribution $p_J$ with mean $\mu_J=1/3$ and standard deviation $\sigma_J=0.1$.
      The initial condition for the dynamics is given by $x_i(0)=1 \, \forall \, i$. The solid
      lines are theoretical results derived from the solutions of  Eq. (\ref{gtwar5}) using the population dynamics algorithm with $N_{\rm pop}=10^5$. The
      symbols are obtained from the numerical solutions of Eq. (\ref{uds}) for $h_i(t)=\xi_{i}(t)=0$ and an ensemble with $50$ independent random networks generated
      from the configuration model with $N$ nodes. The vertical bars are the standard deviations around the mean values. 
      Left panel: dynamics for the exponent $\gamma=3$, such that $c \simeq 3.20$ and the model lies in phase II. Right
      panel: dynamics for the exponent $\gamma=4$, such that $c \simeq 2.45$ and the model lies in phase I.
}
\label{powerlaw}
\end{center}
\end{figure}

\begin{figure}[H]
  \begin{center}
    \includegraphics[scale=1.2]{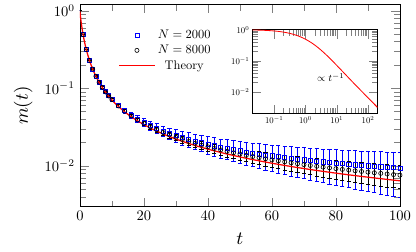}
    \includegraphics[scale=1.2]{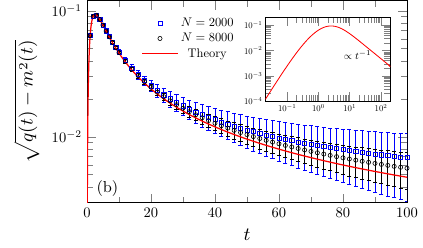}
    \caption{Dynamics of the mean $m(t)$ and the standard deviation $\sqrt{q(t)-m^2(t)}$ of the local dynamical
        variables of the SIS model (see table I in the main text) on directed random networks
      with a Poisson indegree distribution.
      The coupling strengths
      are drawn from an uniform distribution $p_J$ with mean $\mu_J=1/3$ and standard deviation $\sigma_J=0.1$.
      The panels show the relaxation dynamics at the critical mean degree $c=\mu_{J}^{-1}$ that separates the disease-free phase ($m=0$ for $c < \mu_{J}^{-1}$) from
      the endemic phase ($m >0$ for $c > \mu_{J}^{-1}$). The initial condition for the dynamics is given by $x_i(0)=1 \, \forall \, i$. The
      solid red lines are derived from the solutions of Eq. (7) in the main text using the population dynamics algorithm with $N_{\rm pop}=10^5$. The
      symbols are obtained from the numerical solutions of Eq. (\ref{uds}) for $h_i(t)=\xi_{i}(t)=0$ and an ensemble with $50$ independent random networks generated
      from the configuration model with $N$ nodes. The insets show the theoretical results in log-log scale.
}
\label{SIScritic}
\end{center}
\end{figure}

Let us present results derived from the solutions of Eq. (7) in the main text using the algorithm described above. In the main
text, we compare population dynamics with the numerical solutions of the original dynamical equations, Eq. (\ref{uds}), for
heterogeneous networks with Poisson and exponential degrees. To further demonstrate the robustness and accuracy of our theoretical
approach, we extend this comparison to two different scenarios characterized by strong fluctuations: (i) networks with power-law
indegree distributions and (ii) systems prepared at the critical point of a nonequilibrium phase transition. These scenarios are
important tests for our theory, as nonequilibrium phase transitions are a ubiquitous feature of complex systems, and power-law
degree fluctuations are commonly observed in various real-world networks \cite{Newman2010}. Figure  (\ref{powerlaw}) illustrates
the dynamics of the NN model (see table I in the main text) on networks with a power-law indegree distribution, while
Fig. \ref{SIScritic} shows results for the relaxation dynamics of the SIS model at the epidemic threshold. In both cases, despite
the presence of strong fluctuations in the dynamical variables, the numerical solutions of Eq. (\ref{uds}) consistently converge
to the population dynamics results as the system size $N$ increases. These results further support the exactness of Eq. (7) in
the main text for networks with power-law indegrees and for systems at the critical point of a nonequilibrium phase transition.
Note, in addition, that the results in Fig. (\ref{powerlaw}) are consistent with the transition between phases I and II (see Fig. 2 in
the main text). As shown in Fig. \ref{SIScritic}, the theoretical results indicate that the mean and the standard deviation of
the SIS model decay as $m(t) \propto t^{-1}$ ($t \gg 1$) and $\sqrt{q(t)-m^2(t)} \propto t^{-1}$ ($t \gg 1$) at the epidemic threshold.

\begin{figure}[H]
  \begin{center}
    \includegraphics[scale=0.85]{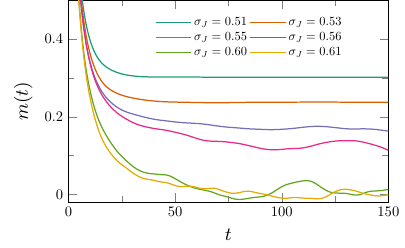}
    \includegraphics[scale=0.85]{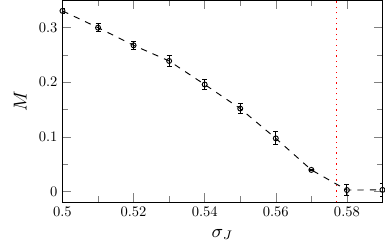}
    \includegraphics[scale=0.85]{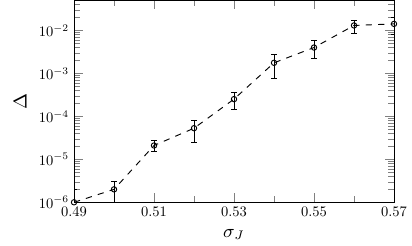}
    \caption{Behaviour of $m(t)$ for the NN model (see table I in the main text) on directed random networks with a Poisson indegree distribution with mean degree $c=4$. The random coupling strengths
      are drawn from a Gaussian distribution $p_J$ with mean $\mu_J=1/3$ and standard deviation $\sigma_J$. The
      results are derived from the solutions of  Eq. (\ref{gtwar5}) using the population dynamics algorithm with $N_{\rm pop} = 5 \times 10^4$.
      The panels illustrate the dynamics and the stationary behaviour of $m(t)$ in phases II, III, and IV (see Fig. 2 in the main text). Left panel: dynamics of $m(t)$ for several values of $\sigma_J$
      and initial condition  $x_i(0)=1 \, \forall \, i$. Middle panel: temporal average $M$ (see Eq. (\ref{M})) of $m(t)$ as a
      function of $\sigma_J$ across the transition between phases III and IV. The vertical dotted
      line marks the gap-gapless transition (see the dashed line in Fig. 2 of the main text). Right panel: temporal standard deviation $\Delta$ (see Eq. (\ref{Delta})) of $m(t)$ across the transition
      between phases II and IV. The results for $M$ and $\Delta$ are averaged over five independent realizations
      of the population dynamics, with the error bars indicating the standard deviation of these realizations. 
%
}
\label{trans}
\end{center}
\end{figure}

In Fig. \ref{trans}, we present results for the dynamics and the stationary behaviour of $m(t)$ in phases II, III and IV of the NN model (see Fig. 2 in the main text). Figure \ref{trans}
  illustrates the behaviour of the temporal mean
\begin{equation}
  M =  \frac{1}{T-T_{\rm tr}}\int_{T_{\rm tr}}^T dt \, m(t)  \label{M}
\end{equation}
and the temporal standard deviation $\Delta$
  \begin{equation}
  \Delta^2 =  \frac{1}{T-T_{\rm tr}}\int_{T_{\rm tr}}^T dt \, \left[ M - m(t)  \right]^2
  \label{Delta}
\end{equation}
  of the macroscopic variable $m(t)$, where $T_{\rm tr}$ is the transient
  time after which $m(t)$ stabilizes in an attractor (either fixed-point or chaotic).
In the limit $T \rightarrow \infty$, the parameters $M$ and $\Delta$ distinguish the different phases of the NN model: $M=0$ and $\Delta=0$ in phase I, $M \neq 0$ and $\Delta=0$
in phase II, $M=0$ and $\Delta > 0$ in phase III, and $M \neq 0$ and $\Delta > 0$ in phase IV.
By numerically solving  Eq. (\ref{gtwar5}), we calculate $M$ and $\Delta$ for $T \gg 1$ and estimate the transition lines that
delimit phase IV. The numerical results for these transitions, shown in Fig. 2 of the main text, are derived
by assuming that a nonzero value of $M=\mathcal{O}(10^{-3})$ identifies the transition from the chaotic phase III to phase IV, while a nonzero $\Delta = \mathcal{O}(10^{-4})$ distinguishes
phase IV from the fixed-point phase II. For $c=4$ and $\mu_J=1/3$, the gap-gapless transition occurs at $\sigma_J = 1/\sqrt{3}$, which is obtained from
the equation $c=1 + \sigma_J^2/\mu_J^2$ (see the dashed line in Fig. 2 of the main text).
The value $\sigma_J = 1/\sqrt{3}$
is marked by the vertical dotted line in the middle panel of Fig. \ref{trans}. Note that this critical value for the gap-gapless
transition is consistent with the vanishing of $M$, which identifies the transition between phases III and IV.
%

\begin{figure}[H]
  \begin{center}
    \includegraphics[scale=1.2]{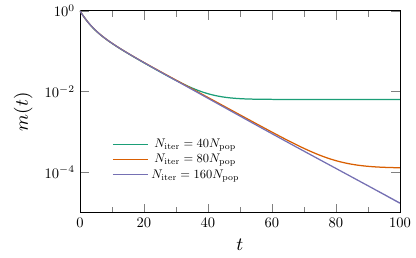}
    \includegraphics[scale=1.2]{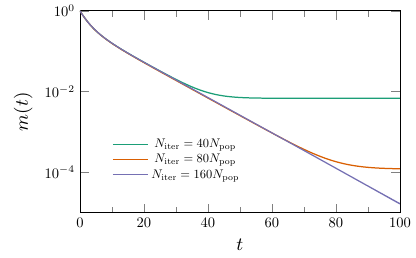}
    \caption{ Dynamics of $m(t)$ for the NN model (see table I in the main text) on directed random networks with a Poisson indegree distribution with mean $c=2.7$. The coupling strengths
      are drawn from a Gaussian distribution $p_J$ with mean $\mu_J=1/3$ and standard deviation $\sigma_J=0.1$. The initial
      condition for the dynamics is given by $x_i(0)=1 \, \forall \, i$. These
      results follow from the solutions of  Eq. (\ref{gtwar5}) using the population dynamics algorithm with population size $N_{\rm pop}$ and total number $N_{\rm iter}$ of
      single-trajectory updates (see the algorithm above). Left panel: $N_{\rm pop}=5 \times 10^{4}$. Right panel: $N_{\rm pop}=10^5$.
}
\label{effect}
\end{center}
\end{figure}

Finally, we present results that illustrate how the number $N_{\rm iter}$ of single-trajectory updates affects the accuracy of the population
 dynamics algorithm. Figure \ref{effect} shows population dynamics results for the dynamics of $m(t)$ inside phase I of the NN model (see Fig. 2 in the main text) for
 different values of $N_{\rm iter}$. We note that, when $N_{\rm iter}$ is not large enough, $m(t)$ evolves to
a nontrivial fixed-point, which is not the correct solution predicted by the stability analysis.
By increasing the total number $N_{\rm iter}$ of single-trajectory updates, we find
that $m(t)$ decays exponentially to $m=0$, which is the correct solution inside phase I.
For the large values of $N_{\rm pop}$ considered in Fig. \ref{effect}, the results are
independent of $N_{\rm pop}$.
%
%


\section{The high-connectivity limit}
\label{high}

An important question is to understand whether the solution of the present class of models converges to the well-known results for
fully-connected models as $c \rightarrow \infty$ \cite{Bunin2017,Sompolinsky1988,Martorell2024}. Put differently, we aim to probe the universality
of fully-connected models in the present context. In a series of recent works \cite{Metz2022,Ferreira2023}, it has been shown that spin models on random networks do not
converge to their fully-connected counterparts as $c \rightarrow \infty$, but the final solution is determined by the empirical distribution
of the degrees rescaled by their mean. In the present context, we also expect that the universality of fully-connected architectures breaks down
as $c \rightarrow \infty$, and the path-probability $\mathcal{P}[x]$ retains information about degree fluctuations
even in the high-connectivity limit.
In this
subsection, we explain  how to take the limit $c \rightarrow \infty$ of Eq. (\ref{gtwar4}) and obtain the form of $\mathcal{P}[x]$ in the high-connectivity
limit. We derive generic analytic results, valid for arbitrary functions $f(x)$ and $g(x,x^{\prime})$ (see Eq. (\ref{uds}) and table I in the main text), which
encapsulates a broad range of important models in the study of complex systems.

By using the functional Fourier transform of the Dirac-$\delta_F$ in Eq. (\ref{gtwar4}), we can integrate over the Gaussian noise $\xi(t)$ and obtain the formal
expression
\begin{equation}
  \mathcal{P}[x] = \int D \hat{x} \, e^{i \int dt \, \hat{x}(t) \left[ \dot{x}(t) + f(x(t))  \right]  - \frac{\sigma^2}{2} \int dt \, \hat{x}^{2}(t)   }
  \sum_{k=0}^{\infty} p_{{\rm in},k} \exp{\left[ k \ln \left( \int D x^{\prime} \mathcal{P}[x^{\prime}] \left\langle e^{- i J \int dt \, \hat{x}(t) g(x(t),x^{\prime}(t) )   }   \right\rangle_{J}  \right) \right]},
  \label{uyt}
\end{equation}
where $\langle \dots \rangle_J$ denotes the average over $J$ with respect to its distribution $p_J$. As we are interested in the high-connectivity limit $c \rightarrow \infty$, it is sensible
to rescale the coupling strengths with $c$ in such a way that $\mathcal{P}[x]$ converges to a finite limit. Therefore, we assume that the first two moments of $J$ are given by
\begin{equation}
\int_{-\infty}^{\infty} dJ \, J \, p_J(J) = \frac{\mu_J}{c} \quad \text{and} \quad \int_{-\infty}^{\infty} dJ \, J^2 \, p_J(J) = \frac{\gamma_{J}^2}{c},
\end{equation}
while higher-order moments of $J$ decay faster than $1/c$. For large $c$, we can expand the logarithm in Eq. (\ref{uyt}) up to $\mathcal{O}(1/c)$, obtaining
\begin{align}
  \ln \left( \int D x^{\prime} \mathcal{P}[x^{\prime}] \left\langle e^{- i J \int dt \, \hat{x}(t) g(x(t),x^{\prime}(t) )   }   \right\rangle_{J}  \right) = &-  \frac{ i \mu_J}{c}
  \int dt \, \hat{x}(t) \int D x^{\prime} \mathcal{P}[x^{\prime}] \, g(x(t),x^{\prime}(t)) \nonumber \\
  &- \frac{\gamma_{J}^2}{2 c} \int dt \, dt^{\prime} \, \hat{x}(t) \hat{x}(t^{\prime})
  \int D x^{\prime} \mathcal{P}[x^{\prime}] \, g(x(t),x^{\prime}(t)) \, g(x(t^{\prime}),x^{\prime}(t^{\prime})).
\end{align}  
The above expression depends on functional averages of one-time and two-time quantities with respect to the path-probability $\mathcal{P}[x^{\prime}]$. By defining
the  conditional macroscopic parameters
\begin{equation}
  M(t|x(t)) = \int D x^{\prime} \mathcal{P}[x^{\prime}] \, g(x(t),x^{\prime}(t)) \quad \text{and} \quad
  C(t,t^{\prime}|x(t),x(t^{\prime})) = \int D x^{\prime} \mathcal{P}[x^{\prime}] \, g(x(t),x^{\prime}(t)) \, g(x(t^{\prime}),x^{\prime}(t^{\prime})),
\end{equation}  
the logarithmic contribution can be written in the compact form
\begin{equation}
  \ln \left( \int D x^{\prime} \mathcal{P}[x^{\prime}] \left\langle e^{- i J \int dt \, \hat{x}(t) g(x(t),x^{\prime}(t) )   }   \right\rangle_{J}  \right) = -  \frac{ i \mu_J}{c}
  \int dt \, \hat{x}(t) M(t|x(t)) - \frac{\gamma_{J}^2}{2 c} \int dt \, dt^{\prime} \, \hat{x}(t) \hat{x}(t^{\prime}) C(t,t^{\prime}|x(t),x(t^{\prime})) .
\end{equation}  

Inserting the above expression back into Eq. (\ref{uyt}) and introducing the distribution of rescaled indegrees \cite{Metz2022,Ferreira2023},
\begin{equation}
  \nu_{\rm in}(\kappa) = \lim_{c \rightarrow \infty} \sum_{k=0}^{\infty} p_{{\rm in},k} \delta\left( \kappa - \frac{k}{c} \right),
  \label{in}
\end{equation}  
we get
\begin{align}
  \mathcal{P}[x] &=  \int\limits_{0}^{\infty}  d \kappa \, \nu_{\rm in}(\kappa)
  \int D \hat{x} \, \exp{\left( i \int dt \, \hat{x}(t) \left[ \dot{x}(t) + f(x(t))  - \mu_J \, \kappa \, M(t|x(t)) \right] \right) } \nonumber \\
    &\times \exp{ \left(   - \frac{1}{2} \int dt \, dt^{\prime} \, \hat{x}(t) \hat{x}(t^{\prime})
      \left[ \sigma^2 \delta(t-t^{\prime}) + \gamma_J^2 \, \kappa \, C(t,t^{\prime}|x(t),x(t^{\prime})) \right] \right) }.
  \label{uyt15}
\end{align}
Finally, we can define the conditional covariance matrix
\begin{equation}
\Delta_{\kappa}(t,t^{\prime}|x(t),x(t^{\prime})) = \sigma^2 \delta(t-t^{\prime}) + \gamma_J^2 \, \kappa \, C(t,t^{\prime}|x(t),x(t^{\prime})) 
\end{equation}  
and rewrite Eq. (\ref{uyt15})  as follows
\begin{equation}
  \mathcal{P}[x] =  \int\limits_{0}^{\infty}  d \kappa \, \nu_{\rm in}(\kappa) \int D \omega \, \mathcal{P}[\omega|\kappa] \,
  \delta_F \Bigg[ \dot{x}(t) + f(x(t)) - \mu_J \, \kappa \, M(t|x(t)) - \omega(t)   \Bigg],
  \label{gttdw}
\end{equation}  
where $\omega(t)$ is a temporally correlated noise with the functional distribution 
\begin{equation}
\mathcal{P}[\omega|\kappa] = \frac{1}{\mathcal{N}_{\omega}} \exp{\left( - \frac{1}{2} \int dt dt^{\prime} \omega(t) \omega(t^{\prime}) \Delta_{\kappa}^{-1}(t,t^{\prime}|x(t),x(t^{\prime}))   \right)  },
\end{equation}  
conditioned to a fixed rescaled degree $\kappa$. The factor $\mathcal{N}_{\omega}$ ensures
that $\int D \omega \mathcal{P}[\omega|\kappa] = 1$.

Equation (\ref{gttdw}) gives the analytic form of the path-probability density in the limit $c \rightarrow \infty$. Clearly, $\mathcal{P}[x]$ is determined
by the distribution $\nu_{\rm in}(\kappa)$ of rescaled indegrees, which means that the effective dynamics retains information about the network
degree fluctuations even in the limit $c \rightarrow \infty$. Put differently, Eq. (\ref{gttdw}) demonstrates that the effective dynamics of fully-connected
networks is not universal, in the
sense that models on sparse networks do not generally flow to their fully-connected counterparts as the mean degree $c$ diverges.
This (perhaps surprising) nonuniversal character is not exclusive to the continuous-time dynamics of complex systems, but it has been previously
identified in the context of Ising spin models \cite{Metz2022,Ferreira2023} and random matrix theory \cite{Metz2020,Silva2022}.

For homogeneous networks, in which the indegree distribution $p_{{\rm in},k}$ is such that $\nu_{\rm in}(\kappa) = \delta(\kappa-1)$, we expect to recover previous
analytic results for the effective dynamics of models on fully-connected networks.
Typical examples of homogeneous networks are
regular networks ($p_{{\rm in},k} = \delta_{k,c}$) and networks with Poisson indegrees ($p_{{\rm in},k} = c^{k} e^{-c}/k!$). In both cases, it is straightforward
to check from  Eq. (\ref{in}) that $\nu_{\rm in}(\kappa) = \delta(\kappa-1)$.
%
Let us explicitly recover previous analytic results in the case of homogeneous networks.
Here we focus on two particular examples: the Lotka-Volterra (LV) model and the neural network (NN) model of \cite{Sompolinsky1988}.
For the NN model, we have $f(x) = x$ and $g(x,x^{\prime}) = \tanh{(x^{\prime})}$, and Eq. (\ref{gttdw}) assumes the form
\begin{align}
\mathcal{P}_{\rm NN}[x] =  \int\limits_{0}^{\infty}  d \kappa \, \nu_{\rm in}(\kappa) \int D \omega \, \mathcal{P}[\omega|\kappa] \,
  \delta_F \Bigg[ \dot{x}(t) + x(t) - \mu_J \, \kappa \, M(t) - \omega(t)   \Bigg],  
  \label{uyts15}
\end{align}
where the covariance of $\omega(t)$ reads
\begin{equation}
\Delta_{\kappa}(t,t^{\prime}) = \sigma^2 \delta(t-t^{\prime}) + \gamma_J^2 \, \kappa \, C(t,t^{\prime}), 
\end{equation}  
with
\begin{equation}
  M(t) = \int D x \mathcal{P}[x] \, \tanh \left( x(t) \right) \qquad \text{and} \qquad
  C(t,t^{\prime}) = \int D x \mathcal{P}[x]  \tanh \left( x(t) \right) \,  \tanh \left( x(t^{\prime}) \right).
  \label{popo}
\end{equation}  
Equations (\ref{uyts15}-\ref{popo}) enable to study the role of indegree heterogeneities  in the high-connectivity limit of the
NN model. As expected, when $\nu_{\rm in}(\kappa) = \delta(\kappa-1)$, Eqs. (\ref{uyts15}-\ref{popo}) reduce to
the equations describing the effective dynamics of fully-connected models with Gaussian interactions \cite{Sompolinsky1988,Crisanti2018,Schuecker2018}.
For the LV model, we have that $f(x) = x(x-1)$ and $g(x,x^{\prime}) = x x^{\prime}$, and the path-probability becomes 
\begin{equation}
  \mathcal{P}_{\rm LV}[x] =  \int\limits_{0}^{\infty}  d \kappa \, \nu_{\rm in}(\kappa) \int D \omega \, \mathcal{P}[\omega|\kappa] \,
  \delta_F \Bigg[ \dot{x}(t) + x(t) \left( x(t)-1 \right) - \mu_J \, \kappa \, x(t) M(t) - \omega(t)   \Bigg],
  \label{g1w}
\end{equation}  
where the covariance of $\omega(t)$ reads
\begin{equation}
\Delta_{\kappa}(t,t^{\prime}|x(t),x(t^{\prime})) = \sigma^2 \delta(t-t^{\prime}) + \gamma_J^2 \, \kappa \,x(t) x(t^{\prime})  C(t,t^{\prime}), 
\end{equation}  
with
\begin{equation}
  M(t) = \int D x \mathcal{P}[x] \, x(t) \qquad \text{and} \qquad
  C(t,t^{\prime}) = \int D x \mathcal{P}[x] x(t) \, x(t^{\prime}).
  \label{ytwq}
\end{equation}  
When $\nu_{\rm in}(\kappa) = \delta(\kappa-1)$ and $\sigma=0$, Eqs. (\ref{g1w}-\ref{ytwq}) yield the effective dynamics
of the corresponding fully-connected Lotka-Volterra model with Gaussian interactions \cite{Galla2018}. The results
obtained here for the Lotka-Volterra model in the limit $c \rightarrow \infty$ are closely related to recent works \cite{Park2024,Aguirre2024}.

Finally, we discuss the correspondence between the phase diagram of the fully-connected version
of the NN model (see Figure 1 in \cite{Mastrogiuseppe2018}) and Fig. 2 in the main text as $c \rightarrow \infty$. The linear stability of the
trivial solution $\boldsymbol{x} = 0$  is governed by the real part of leading eigenvalue $\lambda_1$ of the adjacency matrix $\boldsymbol{A}$.
If ${\rm Re} \lambda_1 < 1$, the trivial solution is stable.
By rescaling the moments of the coupling
strengths in the sparse directed network as $\mu_J = J_0/c$ and $\sigma_J^2 = g^2/c$, the spectrum
of $\boldsymbol{A}$ converges, in the limit $c \rightarrow \infty$, to a uniform disc of radius $|\lambda|=g$ in the complex
plane \cite{Metz2019, Neri2020,Metz2021}. Additionally, when $J_0 > g$, the dense matrix $\boldsymbol{A}$ has an outlier eigenvalue at $\lambda=J_0$.
Thus, combining these results for the leading eigenvalue $\lambda_1$ with the condition ${\rm Re} \lambda_1 < 1$, we find that, in the limit $c \rightarrow \infty$, $\boldsymbol{x} = 0$ is stable
when both $J_0 < 1$ and $g< 1$, while the straight line $J_0=g$ represents the gap-gapless transition.
Hence, the transition lines delimiting the stability of $\boldsymbol{x}=0$ agree with the results
in \cite{Mastrogiuseppe2018}, while the transition to the
homogeneous chaotic phase III in the phase diagram $(J_0,g)$ of \cite{Mastrogiuseppe2018} slightly deviates
from the straight line $J_0=g$ marking the gap-gapless transition in the network spectrum.







\bibliography{bibliography}